\documentstyle[epsf,epsfig,12pt]{article}

\advance\textheight by \topskip
\newcommand{\sla}{\kern -5.4pt /}
\newcommand{\Dir}{\kern -6.4pt\Big{/}}
\newcommand{\Dirin}{\kern -10.4pt\Big{/}\kern 4.4pt}
\newcommand{\DDir}{\kern -7.6pt\Big{/}}
\newcommand{\DGir}{\kern -6.0pt\Big{/}}

\newcommand{\be}{\begin{equation}}
\newcommand{\ee}{\end{equation}}
\newcommand{\bea}{\begin{eqnarray}}
\newcommand{\eea}{\end{eqnarray}}
\newcommand{\beanon}{\begin{eqnarray*}}
\newcommand{\eeanon}{\end{eqnarray*}}
\newcommand{\ba}{\begin{array}}
\newcommand{\ea}{\end{array}}
\newcommand{\bd}{\begin{description}}
\newcommand{\ed}{\end{description}}
\newcommand{\bi}{\begin{itemize}}
\newcommand{\ei}{\end{itemize}}
\newcommand{\ben}{\begin{enumerate}}
\newcommand{\een}{\end{enumerate}}
\newcommand{\bc}{\begin{center}}
\newcommand{\ec}{\end{center}}

\newcommand{\ar}{\rightarrow}
\newcommand{\parno}{\par\noindent}
\newcommand{\vsk}{\vskip 10 pt\noindent}
\newcommand{\hsk}{\hskip 10 pt\noindent}

\newcommand{\wph}{{\tt WPHACT }}

\def\pl #1 #2 #3 {{\it Phys.~Lett.} {\bf#1} (#2) #3}   
\def\np #1 #2 #3 {{\it Nucl.~Phys.} {\bf#1} (#2) #3}
\def\zp #1 #2 #3 {{\it Z.~Phys.} {\bf#1} (#2) #3}
\def\pr #1 #2 #3 {{\it Phys.~Rev.} {\bf#1} (#2) #3}
\def\prep #1 #2 #3 {{\it Phys.~Rep.} {\bf#1} (#2) #3}
\def\prl #1 #2 #3 {{\it Phys.~Rev.~Lett.} {\bf#1} (#2) #3}
\def\intj #1 #2 #3 {{\it Int. J. Mod. Phys.} {\bf#1} (#2) #3}
\def\mpl #1 #2 #3 {{\it Mod.~Phys.~Lett.} {\bf#1} (#2) #3}
\def\rmp #1 #2 #3 {{\it Rev. Mod. Phys.} {\bf#1} (#2) #3}
\def\cpc #1 #2 #3 {{\it Comp. Phys. Commun.} {\bf#1} (#2) #3}
\def\xx #1 #2 #3 {{\bf#1}, (#2) #3}

\begin{document}
\tolerance=100000
\thispagestyle{empty}
\setcounter{page}{0}

\begin{flushright}
{\large DFTT 16/96}\\
{\rm July 1996\hspace*{.5 truecm}}\\
\end{flushright}

\vspace*{\fill}

\bc
{\Large \bf WPHACT 1.0 \\
\vskip 1truecm
A program for $WW$, Higgs and 4 fermion physics at $e^+ e^-$ colliders.
\footnote{ Work supported in part by Ministero
dell' Universit\`a e della Ricerca Scientifica.\\[2 mm]
e-mail: ballestrero,accomando@to.infn.it}}\\[2.cm]
{\large Elena Accomando and Alessandro Ballestrero}\\[.3 cm]
{\it INFN, Sezione di Torino, Italy}\\
{\it and}\\
{Dipartimento di Fisica Teorica, Universit\`a di Torino, Italy}\\
{\it v. Giuria 1, 10125 Torino, Italy.}\\
\ec

\vspace*{\fill}

\begin{abstract}
{\normalsize
\noindent
{\tt WPHACT} ({\bf W} W and Higgs  Physics with {\bf PHACT}) is a MC program
and unweighted event generator which computes all Standard Model processes 
with four 
fermion in the final state at $e^+ e^-$ colliders. 
It is based on an helicity amplitude method which allows precise
and fast evaluations of the matrix elements both for massless and massive
fermions.  Fermion masses for $b$ quarks are exactly 
taken into account. QED initial state and Coulomb
corrections are  evaluated, while QCD final state corrections are included in
an approximate formulation. 
Cuts can be easily introduced and distributions for any variable at parton
level can be implemented. The contributions to the processes of neutral
Standard Model or Susy Higgs can be included. 
Anomalous couplings effects for the triple coupling can be computed. 
An interface to hadronization is provided and Jetset can be directly called 
from the program. }
\end{abstract}

\vspace*{\fill}
\newpage
\section*{Program Summary}
\vskip 15pt

\leftline{{\it Title of program:} {\tt WPHACT}}
\vskip 8pt

\noindent
{\it Program obtainable by:} anonymous ftp from ftp.to.infn.it in the
 directory pub/ballestrero.
\vskip 8pt

\noindent
{\it Computer:} DEC VAX, DEC ALPHA AXP, HP/APOLLO;
{\it Installation:} INFN, Sezione di Torino, via P.~Giuria 1, 10125 Torino,
Italy
\vskip 8pt

\leftline{{\it Operating system:} VMS, OVMS, UNIX}
\vskip 8pt

\leftline{{\it Programming language used:} FORTRAN 77}
\vskip 8pt

\noindent
\leftline{{\it Memory required to execute with typical data:} $\approx$ 
500 KByte}
\noindent

\vskip 8pt

\leftline{{\it No. of bits in a word: } 32}
\vskip 8pt

\noindent
{\it Subprograms used:} {\tt VEGAS}\cite{vegas},
{\tt GAMMLN}\cite{numrec}, {\tt RAN2}\cite{numrec}.
\vskip 8pt

\leftline{{\it No. of lines in distributed program:} $\approx$ 15600 }
\vskip 8pt

\noindent
{\it Keywords:} high energy electron-positron collisions,
four-fermion final state, $W$-pair production, Higgs, $Z$-pair production,
LEP2, NLC, QED corrections, electron structure functions, Coulomb corrections,
anomalous couplings.
\vskip 8 pt

\leftline{{\it Nature of physical problem} }
\noindent
The forthcoming experiments at the high energy electron--positron
collider LEP2 will be mainly concerned with $WW$ physics and Higgs search.
The production of two $W$'s will allow the direct study of the triple-boson
coupling and a precise measurement of the mass of the $W$. The search for
the Higgs is of primary importance for  understanding the problem of mass
generation in the Standard Model (SM). Small deviations from the SM will be
important for discovering possible new physics. Both $WW$ and Higgs production
 will result in a four-fermion final state. It is therefore mandatory to have
accurate predictions for all physical processes with a four fermion final state
in order to have full control on signals as well as backgrounds to the
processes of interest. The same kind of processes will also play a fundamental
role in electron--positron accelerators at higher energy which will be able
to extend Higgs search at higher values of the mass and probe triple gauge
boson physics and gauge cancellations.

\noindent
\vskip 8pt
\leftline{{\it Method of solution} }
\noindent
Full tree level matrix elements for all  processes are computed
by means of subroutines which make use of the helicity formalism of ref.
\cite{method}-\cite{phact}. The number of Feynman diagrams in the various 
channels varies
from $3$ to $144$. The velocity in computing these amplitudes that the above
mentioned method allows, becomes therefore essential to take exactly into
account fermion masses and to obtain high precision in a reasonable amount of
CPU time.
 \par Different integration variables for the phase space are employed in
order to dispose of the peak structure of the  resonating diagrams for
the different processes. The adaptive routine  {\tt VEGAS}\cite{vegas} is used
 for
the numerical evaluation of the integrals.
 \par Distributions can be produced to study the behaviour of any variable of
interest. For simulation purposes, the program can also be used as an event
generator that provides unweighted events. An interface to standard 
hadronization packages and specifically to Jetset\cite{jetset} is provided.

\vskip 8pt
\leftline{{\it Restrictions on the complexity of the problem} }
\noindent
 Only for processes with $b$'s in the final state the masses of the fermions
are accounted for. Final state radiation is not  implemented. Initial state
radiation (ISR) is included through Structure Functions and no photons $p_t$ is
computed. QCD corrections are introduced only in an approximate way.

\vskip 8pt
\leftline{\it Typical running time}
\noindent

The running time strongly depends on the process considered and on the
precision requested. Some examples are reported in Table 1.
For the typical final state $\mu^-$~$\bar\nu_\mu$~$u$~$\bar d$ with ISR 
the time per call on AlphaStation 600 5/333 is  $6.\times 10^{-5}$ sec.
The longest time for call is $6. \times 10^{-4}$ sec.   for $b$ $\bar b$
$b$ $\bar b$. 
At Lep2 energies, 5 M  calls (about 5 minutes)  are used
    to obtain for $\mu^-$ $\bar\nu_\mu$ $u$ $\bar d$ 
 a cross section (with ISR) with a typical
   estimated relative error of  $2 \times 10^{-4}$ sec.  The same process can be
   evaluated in about $40$ sec.  with less than 1 M calls at permill level.
All above running times have to be multiplied approximately by a factor $3$ 
for an AlphaServer 2100 4/200 computer.
\noindent

\vskip 8pt

\leftline{{\it Unusual features of the program:} }
\noindent
{\tt REAL*8} and {\tt COMPLEX*16} variables and {\tt STRUCTURE} 
declarations are used.
Compilation on Hp/Apollo stations has to be performed with -k option.
\eject
\section*{Long Write-Up}

\section{ Introduction }

After  LEP 1 precision measurements and  top discovery at Tevatron, which have
shown an impressive agreement of the Standard Model (SM) with data, 
a new era is starting  in which the mechanism of spontaneous symmetry breaking
and the non abelian structure of the model will be directly tested 
by the experiments. At LEP 2 it will be possible to measure the contribution
of trilinear gauge boson couplings, and  for the first time the production
of two $W$'s and two $Z$'s will be seen. The properties of the $W$'s 
will be measured
with great accuracy and these will also contribute to put more stringent limits
on the Higgs mass.  At the same time, direct searches for the Higgs boson
will allow to find it, if its mass is not greater than $\approx 100$ GeV. These
studies and the search for the Higgs will be extended to LHC and to the next
$e^+ e^-$ collider which will presumably approach the TeV range. Theoretical
arguments claim for the onset of a 'new physics' regime at this scale, so that
LEP 2 and the future machines will search not only for new particles, like
the MSSM Higgs, but for any deviation from the SM predictions.\par
It is evident in this scenario that theoretical predictions must reach a high
accuracy to confront and analyze the data. $WW$ and Higgs physics and their
radiative corrections have been thoroughly studied in the past (for a complete
review on this subject see ref.\cite{yr}). It has however been realized that
on shell predictions may not be accurate enough, as the measured final states
will not correspond to, say, two $W$'s or a Higgs and a $Z$, but rather to
4 fermions. This implies that one has to deal with irreducible backgrounds, i.e.
contributions to the matrix element for the envisaged final state which do 
not correspond for instance to two $W$'s production and decay and cannot be 
separated from it, even if they can be reduced with appropriate cuts. Moreover
many different  final states are  not experimentally
distinguishable, and one has  to take into
account  that the properties of the $W$ and Higgs bosons can be only 
reconstructed by a careful analysis of all four fermion final states.
For top physics, trilinear and quadrilinear coupling studies and for the Higgs,
if it  will be as heavy as to decay into two $W$'s, also six fermion final
states will have to be analyzed at future $e^+ e^-$ colliders. \par
Several codes for four fermion physics have been produced in the last years
\cite{fort}-\cite{gp}-\cite{wweg}-\cite{dpeg}, and they have been used for
phenomenological studies mainly concerning $WW$ and Higgs Physics 
\cite{phen}-\cite{dima}-\cite{noi}.  Only some of them can produce accurate 
results for all four fermion processes, with the inclusion of all the relative
Feynman diagrams. During last year's LEP~2 Workshop the results produced by
the different codes have been extensively 
compared~\cite{wweg}-\cite{dpeg}-\cite{smwg}. 
The codes have very different characteristics. Some of them are 
classified (see ref.\cite{wweg}) as semi-analytical \cite{dima} 
or as deterministic\cite{gp}. All the others belong to the broad 
Monte Carlo's (MC) class, where a further subdivision can be made among
unweighted event generators and Monte Carlo integration programs.
The semi-analytical codes
 perform as much as possible of the integrations analytically, 
leaving only low dimensional integration to be performed numerically. The MC
and deterministic programs on the contrary perform the whole of the
9-dimensional integration (including ISR) numerically. This implies that by 
their own nature
the semi-analytical codes may easily  reach an extreme precision, but they 
cannot
implement all cuts: normally only those on the invariant masses are viable.
The MC and Deterministic programs can implement all cuts, but of course they
are normally slower and less precise. The deterministic program\cite{gp} 
implements all cuts analitically, while only few of them can be implemented
in MC's as limits on the integration variables.\par
\wph belongs to the family of the MC integrators and event generators.
It has been developed only in the last year but it has been compared and 
tested continuously both in LEP~2 Workshop~\cite{wweg}-\cite{dpeg}-\cite{smwg}
 and in phenomenological studies\cite{noi}. One of the main characteristic
of the code is that of using a new helicity amplitude 
formalism\cite{method}-\cite{phact} which
allows to compute matrix elements in a very fast and precise way. 
As a consequence, the code can reach high precision in a 
relatively short time. This is particularly useful for example when one wants 
to use \wph to produce distributions at parton level. If one 
requests that  the range of a variable be divided
in a large number of bins, of the order of 100 say, one can reach very low 
errors on each bin only producing a number of weighted events of the order of
ten millions.
And this is precisely what \wph can do in a reasonable time (see Table 1)
even for the most complicated processes as $e^+ e^- \ar b \bar b b \bar b$
with all diagrams and taking $b$ masses into account.\par
As already mentioned in the program summary, \wph allows to compute 
QED initial state and Coulomb corrections, and  also QCD final state 
corrections in
a 'naive' formulation, exact  in the limit in which only double 
resonant diagrams are considered and  no cuts are imposed. 
Possible cuts are explicitly provided and any distribution  at parton
level can be implemented. 
\wph is also an unweighted event generator. The events  can be produced while
evaluating the cross section and also after this has been computed.
In this second case any number of events can a priori be requested.
 An interface
to hadronization is in the code and it may be linked directly to Jetset.
As far as 'new physics' is concerned, Anomalous Couplings effects for the 
triple coupling and cross sections for neutral MSSM  Higgs $h$ and $A$ can be 
computed.\par
In presence of unstable gauge bosons the imaginary
part of their propagators violates gauge invariance in tree level computations
of processes like those we are considering. The way to restore it
and get a reliable result is that of including relevant parts of fermionic 
corrections, thus  fulfilling Ward identities\cite{bhf}. 
From a numerical point of view these corrections become relevant only for
some particular processes and cuts. For instance in the process 
$e^-$ $\bar\nu_e$ $u$ $\bar d$ no discrepancy between approximate and correct
computation is appreciable if a cut of the order or greater than $\approx 5^o$
is applied to the
angle between the electron and the beam\cite{noi}.  
The present version of \wph does not include these corrections.
Also final state radiation and effects due to transverse momenta
of QED radiation are not computed at present by {\tt 
WPHACT}.
Subtle theoretical problems are connected to these issues and probably 
only a full computation $O(\alpha)$ can assess them. As far as final radiation
 is concerned, this can in any case  be introduced via Jetset when \wph is used 
as an event generator. \par
In the following we will describe the main features of the code and the way
to make use of it. Some useful examples of test runs will be given at the
end.

\section{ General Features }

\subsection{ Processes }
 \wph computes all SM processes with four fermion in the final state
  at $e^+ e^-$ colliders. Final states with $t$ quarks
  are not considered, as the $t$'s are known to decay immediately to 3
  other fermions.

  The processes are enumerated in Table 2 and 3. They are divided in 4 classes. 
  The first (CC) and the third (NC) contain
  all processes which  have only diagrams with   charged 
 or neutral currents respectively. In the second (Mix) they have both
 kind of diagrams. In the last class (NC+Higgs) we have included the processes
 with $b$'s in the final state, where we may have important contributions from
 the Higgs. These are the only processes in which \wph accounts for
 Higgs diagrams and it treats the $b$ quarks as massive both in the phase space
 and in the matrix elements.

  In \wph the momenta of the particles in the final state, as reported in the
  tables, are respectively
  $p3$, $p4$, $p5$, $p6$. The order of the final particles is important
  when one wants to choose a certain set of cuts for the final state, or
  to ask the program to compute some distribution at parton level.
  It has to be noticed that for CC and Mixed processes the order in which the 
   momenta
  are passed to the high energy standard {\tt COMMON/HEPEVT}\cite{tb} and hence
  also to Jetset may be different from the one used by the program: in the 
  common particles 3 and 4 are respectively the particle and the antiparticle 
  that correspond to a $W^+$, 5 and 6 to a $W^-$.

 Many different final states give the same cross section at  parton level.
 This is the case for instance of  $\mu^-$  $\mu^+$  $d$  $\bar d$ and
 $\tau^-$  $\tau^+$  $s$  $\bar s$ if the  mass differences are neglected.
 For this reason to each final state two flags are attributed, {\tt iproc} 
 and {\tt ich}, which serve to identify it. The first refers to the
group which has the same diagrams (whose number is indicated in the first
column) and cross section, the second to the specific final state of the 
group. It has to be noticed that for CC processes charge conjugate final
states belong to the same {\tt iproc}. The
amplitudes for a CC final state and its charge conjugate for a given set
of four momenta, are not equal.  To obtain the same amplitude one
must consider the parity transformed final state of the charge conjugate.
This implies that if the cuts are not invariant under parity transformation
(i.e. both a set of final four momenta and its parity transformed are accepted
or not accepted ) the two cross sections will be slightly different.
An example of this is given by  $e^+ e^- \ar \mu \bar \nu_\mu u \bar d$ and
$e^+ e^- \ar \mu^+ \nu_\mu \bar u  d$ when in both processes the cut angle
of the $\mu$ with $e^+$ is different from that with $e^-$. \wph accounts
exactly also for these slight differences. The main reason  to
specify the exact final state instead of just indicating the group is
however to distinguish among final states when hadronization via Jetset is 
performed, as this depends on the final state particles.

  The production of two $W$'s which decay in two fermions each is described
  by $3$ Feynman diagrams. The cross section corresponding to these
  $3$ diagrams only is often referred to as CC3\cite{wweg}.
  Of course CC3 does not correspond to physical processes, and the three
  diagrams alone do not constitute a gauge invariant set,
  but for some energies and final states and cuts they might be a reasonable 
  approximation.
  For all CC and CC+NC final states, {\tt WPHACT} can compute the complete
  process or the CC3 contribution to it.

  All neutral current processes are normally computed at order $ \alpha ^4$.
  When there are four quarks in the final state, there are however 
  contributions $O(\alpha ^2 \alpha_s ^2)$ of diagrams in which a gluon is  
  exchanged 
  between the two quark lines. This contribution is of course enhanced
  by the coupling and can be relevant for some processes, energies and cuts. 
  For instance we have found\cite{dpeg} that for
  $\sigma(e^+ e^- \ar b \bar b b \bar b)$ at $E_{cm} = 175$ GeV and with
  $m_{b\bar b}\geq 20$ GeV, it is more than twice the pure electroweak 
  contribution,  while it becomes only about one third at $E_{cm} = 192$ GeV. 
  \wph may or not include this contribution. If one includes it, one must
  be aware that part of it may also be accounted for by two quarks final state
  computations if parton shower and hadronization are added.
  The interplay among these two ways of treating such a
  contribution is surely strongly dependent on cuts used and deserves 
  a careful analysis case by case.
  
\subsection{Helicity amplitudes}
All amplitudes for the processes computed by \wph are evaluated with the
helicity formalism of ref.~\cite{method}, which is based
on the insertion in spinor lines of a complete set of states for every
fermion propagator. These states are  
eigenstates of $p\sla$, where $p$
is the momentum flowing in the fermion propagator. They are 
chosen to be generalizations of the spinors used in ref.\cite{ks}.
In this way one needs not to decompose every $p\sla$ in terms of the external
momenta, as it is the case for some other helicity method, and the 
numerator of the fermion propagator assumes a very simple expression.
The computation is in this way reduced to the evaluation of one  $\tau$
matrix for every vertex and to combine them together. For the definition
of these $4 \times 4$  matrices we refer again to ref.~\cite{method}.
In this base the matrix corresponding to the numerator of every
fermion propagator is diagonalized for massless lines and for massive
lines the non zero off diagonal elements are simply given by the mass
itself. It has to be noticed that a $\tau$ matrix fully describes the
vertex both in the case of insertion of an external particle and in that  of
the insertion  of a whole piece of
diagram. Moreover, combining together two $\tau$ matrices 
corresponding to two vertices,
one obtains a new $\tau$ matrix corresponding to the two vertices and so on.
These simple facts allow to achieve a great modularity in the computation, 
to keep track
of partial results and to set up a recursive scheme which computes and stores
for later use subdiagrams of increasing size and complexity.
Moreover  the massive
case is not more complicated than the massless one. Only more
helicity indices are of course needed. As a consequence, the codes for massive
 amplitudes  written in this way are not much slower, as it is normally the
 case, than those with massless fermions.

 The code for \wph amplitudes has been completely written 
with the help of PHACT \cite{phact} ({\bf P}rogram for {\bf H}elicity
{\bf A}mplitudes
  {\bf C}alculations with {\bf T}au matrices).
This program  implements the  method  described above in a fast and efficient 
way. It essentially writes directly the optimized fortran code for every
$\tau$ matrix corresponding to a vertex and for combining $\tau$ matrices
together. It also writes the code for external gauge boson polarization
vectors, triple couplings, and so on. 
With it, one  avoids as much
as possible computing expressions which will turn out to be zero.
Moreover in the computations of the amplitudes we have  avoided
 any call to external 
subroutines and functions which might be time consuming for such a repetitive
part of the program. This
of course  leads to a somewhat longer program than usual.

\subsection {Phase space and integration variables}\label{ph}
To describe the phase space, the four final fermions are divided in two couples.
Every couple corresponds to two particles that can decay from a $W$ in CC,
from a $Z$ or a Higgs in NC contributions.
Take for instance the case in which $f_3$  $f_4$ and $f_5$  $f_6$ are grouped
together.
 Natural variables are then the two invariant masses $m_{34}=\sqrt{(p3+p4)^2}$,
$m_{56}=\sqrt{(p5+p6)^2}$ and the decay angles of one particle for each
couple  $\theta_3^*, \phi_3^*, \theta_5^*, \phi_5^*$ in the rest frame of the
couple itself. 
\par Whenever the energy is such that a couple of particles can have an
invariant mass $m$ equal to $M$, the mass  of a real $W$ or  $Z$ or  Higgs, the
corresponding amplitude squared will be almost proportional to a Breit-Wigner 
peak:
\be
\frac{1}{\left(m^2-M^2\right)^2+\gamma^2}.
\ee
It is therefore convenient
to use instead of the invariant mass $m$ of the couple an
integration variable proportional to :
\be\label{tang}
x=arctg \frac{M^2-m^2}{\gamma}.
\ee
Every variable is always transformed to the interval $0 \div 1$.
When  substitution (\ref{tang}) is performed on both invariant masses, we refer 
to it  as double resonant mapping. If only one or no substitution is performed 
we have respectively a single resonant or non resonant mapping. 
In \wph there is the possibility to choose among these mappings. Most of the 
results
at LEP 2 energies are better obtained with the double resonant phase space.
For some processes one might have a peak also for low invariant masses, due
to photon propagator contributions. Whether or not this peak becomes
relevant depends strongly on the cuts. When such is the case, one has a 
variable $m$ to which there correspond
two peaks: the Breit-Wigner and the photonic one. After having tried different
mappings, we have found that, given the fact that {\tt VEGAS} is an adaptive 
routine,
it is better just to use a non resonant mapping for $m$ in this case.

When one has to deal with a Mixed (or NC+Higgs) process, the peaking structure
of CC (Higgs) and that of NC contributions is different. In such cases
\wph automatically integrates separately the two contributions. To one of them
it is added the interference.

The contribution to $e^+ e^- \ar b \bar b b \bar b$ of diagrams with  
$h$ $A$ intermediate states and that of $h$ $Z$   give two ($A$ and $Z$) 
Breit Wigner
peaks on the same variable $m$. In this case a double mapping of the type 
(\ref{tang}) is performed simultaneously on it.

With ISR two more integrations are to be performed and we use also 
for this case a change of variables to level the form of the distribution 
functions, as we will see in the following section.

All integrals are computed with {\tt VEGAS}.  For  this routine
it is convenient  to use normally more than one iteration. In every iteration
the integral is evaluated and at the end the various results are combined 
together. This allows to optimize  the number of points computed in the various
regions of the integration variables. It may also be
useful to perform some thermalizing iteration with  a lower number
of points to be evaluated. The thermalizing iterations are just used
to adjust the grid and not for the final result. 

The adaptivity of {\tt VEGAS} is such that, even if most cuts are  
implemented in the program with the use of {\tt if} statements
which act as $\theta$
functions, usually this does not
correspond to a sensible lost in time and precision.
 The cuts on the invariant masses which are function of integration  variables
are implemented directly on integration limits. 

\subsection {QED and QCD corrections}

QED initial state radiation is taken into account in \wph via the structure
function approach\cite{sf}. In it, the  cross section $\sigma (s)$ is obtained
by means of  a convolution  with functions $D(x,s)$ which account for the 
radiative emission by the initial particles of a fraction $x$ of their energy:
\be
\sigma (s) = \int d x_1 \, d x_2 \, D(x_1,s) D(x_2,s) \, \sigma (x_1 x_2 s)
\ee
The expression for the $O(\alpha^2)$\, $D(x,s)$ in leading log approximation
used by \wph is: 

\bea
D(x,s)&=&\, { {\exp \left[ \left( \frac{3}{8} -\frac{\gamma_E}{2}
\right)\beta\right] }
\over {2 \,\Gamma \left( 1 + \frac{1}{2} \beta \right) } } \,
\beta (1 - x)^{\beta/2 - 1} - {{\beta} \over 4} (1+x) \\ \nonumber
&+&- {1 \over {32}} \beta^2 \left[ 4 (1+x) \ln(1-x)  +
 {1+3x^2 \over {1-x}}\ln x + 5 + x \right] ,
\eea

\noindent
with $\beta = 2\,\alpha / \pi\,(\ln ( {s / {m_e^2}})-1)$
and                                                            
$\gamma_E$  the Euler constant.
\par
Given this form of structure functions, it is convenient for the MC to
perform a change of variables in order to flatten out the peak of the
distributions near $x=1$. Thus we have chosen 
instead of $x_1$ and $x_2$ the integration variables
\be
y_i=(1-x_i)^{\beta/2}  \quad\quad (i=1,2)      
\ee
Another important QED
correction to be accounted for is  the so called Coulomb
singularity,
which is due to the electromagnetic interaction of the two $W$ bosons at low
velocity. It regards of course only CC and Mixed processes near the threshold 
energy for production of two $W$'s. In that region  however its
correction amounts to a few percent. In \wph  this correction is computed
with the method of ref.\cite{coul}.

Final state QCD corrections are taken into account by means of the
so called 'naive' approach. In it, the various contributions to the amplitude
are multiplied by the corrections relative to the decay width 
of the $W$'s, of the $Z$'s or
of the Higgs. This approach disregards all 
corrections to non resonant diagrams and corrects exactly  the decay vertices
only if no cuts are applied. Nevertheless they represent at present an estimate
of a sizeable effect. \par
    For NC, CC and Mixed processes the QCD corrections
    appear in an overall factor:
    $(1+{\alpha_s\over{\pi}})$ for semi-leptonic case and
    $(1+2{\alpha_s\over{\pi}})$ for non-leptonic one.
    Gluon exchange diagrams and $CC(NC)\otimes QCD$ interference are not 
    multiplied by a correction factor as we only consider the first order in 
    $\alpha_s$.
    For the processes involving $b$ quarks, Higgs exchange diagrams receive the
    following strong correction:
    $(1+5.67{\alpha_s\over{\pi}})$ for semi-leptonic case and
    $(1+6.67{\alpha_s\over{\pi}})$ for non-leptonic one.
    The $NC\otimes Higgs$ interference is multiplied by
    $(1+3.335{\alpha_s\over{\pi}})$ for semi-leptonic case and
    $(1+4.335{\alpha_s\over{\pi}})$ for non-leptonic one.
By default  $\alpha_s$ in the preceding formulas is taken at $M_W$ scale
for CC diagrams, at $M_Z$ scale for NC and NC+Higgs ones.
Its value can however be changed in a {\tt DATA} statement.
For consistency, when corrections are applied to the vertices they should also
be present in the same way in the widths appearing in the propagators of the
bosons.
This is surely achieved
if one chooses  to let \wph compute the corresponding total widths.

\subsection{Susy and Anomalous couplings}
Besides SM Higgs processes in two $b$'s and two other fermions, \wph computes
also Susy neutral Higgs production in the same final channels. The only
MSSM process which cannot be deduced just changing coupling constants with 
respect
to the SM is $e^+e^-\ar h A \ar b \bar b b \bar b$, where $h$ is the lightest
CP-even and $A$ the CP-odd Susy Higgs. The cross section for the above
diagram is computed as well as all other contributions to 4 $b$'s
final state. As for all other Higgs processes, the contributions of all
diagrams containing the Higgses can be optionally separated from the rest.
The Susy parameters to be given in input  are  
the mass of the pseudoscalar Higgs $A$,
and the ratio between the two vacuum expectation
values $tg(\beta)$. In the simplest version of the  MSSM, all Higgs 
masses are predicted in terms of these two parameters.
At one--loop these predictions are substantially modified and an additional
dependence on the top mass $m_t$ and on the common squark mass
$m_{\tilde t}$ is introduced.
We have used the following relations:
\begin{eqnarray}
M^{2}_{h}& = & \frac{1}{2} [ M^{2}_{A} + M_{Z}^{2} +
\epsilon/\sin^{2}\beta ]\nonumber \\
           &   & - \frac{1}{2} \left\{ [ (M^{2}_{A} - M^{2}_{Z})\cos2\beta +
\epsilon/\sin^{2}\beta]^{2}
                 +(M^{2}_{A} + M^{2}_{Z})^{2}{\rm sin}^{2}2\beta
\right\}^{1/2} ,
\end{eqnarray}
where
\begin{equation}
\epsilon = \frac{3e^{2}}{8\pi^{2} M^{2}_{W}{\rm sin}^2\theta_W}m_{t}^{4} {\rm
ln}\left( 1 +
\frac{{m}^{2}_{\tilde t}}{m_{t}^{2}} \right).
\end{equation}
The squark mass scale ${m}_{\tilde t}$ has been chosen to be 1 TeV.
The mixing angle $\alpha$ in the $CP$--even sector, which together with
$\beta$ determines all couplings of the  MSSM Higgses,
is defined by
\begin{equation}\label{m3}
\tan 2\alpha = \frac{(M_{A}^{2} + M_{Z}^{2}){\rm sin}2\beta}{(M_{A}^{2} -
M_{Z}^{2})
{\rm cos2}\beta + \epsilon/{\rm sin}^{2}\beta}.
\end{equation}

As far as Anomalous Couplings computations are concerned, we have implemented
those relative to the trilinear vertex. 
Starting from the most general effective lagrangian\cite{efflagr}, 
one gets for them nine possible couplings   
just imposing Lorentz and electromagnetic invariance. These are further reduced
to $6$ if $CP$ invariance is imposed, and $5$ of them are separately $C$ and
$P$ invariant. Among the various possible parametrizations of these $6$
quantities, we have implemented that of ref\cite{ancoup}. 

\subsection{Distributions and unweighted event generation}
One of the main purposes of a dedicated code as \wph is to perform 
phenomenological studies and confront theoretical predictions with experiments.
To this aim,  the possibility of computing 
differential cross sections or distributions is extremely important. 
Special care has been devoted
to this aspect and practically any distribution at parton 
level can be computed while the total cross section is evaluated.
Explanations of how to request and implement distributions are given
in section 3.1. In practice, one has just to write down in an include
file the definition of the variables to be distributed in terms of the
4-momenta of the outgoing particles and to specify in input the number
of bins and the interval for every such variable. Files with the extension
.dat  will contain in output the cross section and the evaluated statistical 
error relative to every single bin. A large amount of bins (of the order of
 100) per variable  with a low statistical error can easily be achieved.

If the distribution refers to leptons in
the final state, it will correspond to some directly measurable variable.
For quarks, the hadronization process might prevent this possibility.
Distributions at parton level are however much faster to obtain
than the ones with particles which have to be computed after parton shower 
and  hadronization.
They constitute a very effective tool
to study the physical problem at hand and to deconvolute perturbative effects 
from non perturbative ones. Examples of two distributions obtained with
\wph are given in figs. 1,2.

With \wph it is also possible to generate unweighted events and to store
their 4-momenta. From them one can of course successively produce any
distribution. To reach low statistical errors in this way a high number
of events has to be stored, but they can be produced in a reasonable time:
for instance after having evaluated the cross section, 
it takes $48$ min. to produce 500000 unweighted events for MIX19 
$e^+ e^- \ar \mu^-$  $\mu^+$  $\nu_\mu$  $\bar\nu_\mu$.

 \wph can generate
unweighted events with the hit or miss method while evaluating the
integral. In that case the first effective iteration after thermalization
is used to find the maximum and the second to generate unweighted events.
 One may also produce them
after having evaluated the integral. This feature can be used both
to readjust the maximum if some event exceeded it with the first method,
or to ask for a  predetermined number of events to generate.
When unweighted events are produced, they may be passed directly to 
Jetset using the routine {\tt AB\_LU4FRM}.  So one can for instance run the 
first time just to find an appropriate
  maximum and a correct cross section (and eventually to produce parton
  distributions). Afterwards one verifies  whether it is convenient to 
  multiply the  maximum for an appropriate factor and with a second run 
  one  generates  the desired number of events. To save time, the
  parton shower and hadronization   procedure  with the link to JETSET may be 
   activated only in   this second run. 

 As explained in sect. \ref{ph} , the integration of Mixed and NC+Higgs 
  complete processes is 
  performed in two steps, with different mappings of the phase space. 
 Also the event generation will be performed in two steps  in these
cases.  This implies that the events that come first will have been  
 selected  with the CC (Higgs) contribution and the remaining with the NC 
 contribution.  The resulting sample will have the right proportion of 
 events, but from  them one cannot take away an arbitrary part. If for some 
 reason  it is necessary to diminish the generated events and still have
 an umbiased sample for the whole process, this must be done 
 choosing at random which events are to be taken.

\section{ Program Structure }

In the following two sections we will explain how to use the input
parameters to exploit the various possibilities of {\tt WPHACT}, and the 
meaning
of the various subroutines and functions of the code. At the end of the paper
we report and briefly comment some significant examples of test runs.

\subsection{Input} \label{input}
The way the input parameters must be given in the command file is easily 
understood just looking at the program lines containing the {\tt READ} 
statements.
We therefore reproduce them and explain their meaning in the following.
\par
Every parameter  whose initial is {\tt i} or {\tt n} is
{\tt integer*4}. All  others are {\tt real*8}.  
When a variable has a yes/no option the value {\tt 1} corresponds to {\tt YES},
 {\tt 0} to {\tt NO}.
All energies must be expressed in GeV.
\vsk
{\tt READ*,e\_cm}\parno
{\tt e\_cm} is the centre of mass energy. 
\vsk
{\tt READ*,iproc}\parno     
{\tt READ*,ich}\parno      
The value for  these two parameters must be read directly from Table 2 
and 3. {\tt iproc}  selects the group of processes with the same cross
section at parton level with symmetric cuts, while {\tt ich} selects among 
them the specific final state. The 
choice of {\tt ich} is relevant only if hadronization is performed via
Jetset or if in CC processes cuts are considered for which a final 
configuration
and its charge conjugate behave differently. In all other cases {\tt ich=1} 
gives the same results as any other value indicated in the tables.
\vsk
{\tt IF(iproc.GE.33)THEN}\parno
{\tt  \hsk   READ*,rmb       }\parno
{\tt  \hsk  READ*,icch      }\parno
{\tt \hsk IF(icch.NE.2)THEN}\parno	
{\tt \hsk\hsk  READ*,isusy}\parno
{\tt \hsk\hsk  IF(isusy.EQ.0)THEN}\parno
{\tt \hsk\hsk\hsk    READ*,rmh       }\parno
{\tt \hsk\hsk  ELSE}\parno
{\tt   \hsk\hsk\hsk  READ*,rma       }\parno
{\tt   \hsk\hsk\hsk  READ*,tgb       }\parno
{\tt \hsk\hsk  ENDIF}\parno
{\tt \hsk ENDIF}\parno
{\tt ENDIF}\parno
The processes for which {\tt iproc $\geq$ 33} are those with massive $b$'s 
in the final state. {\tt rmb} fixes the $b$ mass for the phase
space and the matrix elements. The $b$ mass in the Higgs
   coupling  may be different from {\tt rmb} and it has to be set in the 
{\tt DATA}.
 In these processes one can have diagrams with
SM  or Susy MSSM  neutral ($A$ or $h$) Higgs. One has the possibility
of choosing to compute only these diagrams 
(Higgs signal: {\tt icch=1}), only  those without Higgs (to which we refer
as Higgs background: {\tt icch=2}) or the whole set of diagrams 
(Higgs+Background+interference: {\tt icch=3}).\parno
{\tt isusy=0} corresponds to SM Higgs, {\tt isusy=1} to Susy Higgs.
In the first case one has to specify the Higgs mass ({\tt rmh}), in the second
the mass of the pseudoscalar Higgs A ({\tt rma}),
and the ratio between the two vacuum expectation
values $tg(\beta)$ ({\tt tgb}).
\vsk
{\tt IF (iproc.GE.6.AND.iproc.LE.8) THEN    }\parno
{\tt  \hsk READ*,iccnc     }\parno
{\tt ENDIF}\parno
The processes for which {\tt 6 $\leq$ iproc $\leq$ 8} are those which have
both CC and NC diagrams (Mixed).  For these  one may choose to compute
only the CC contribution ({\tt iccnc=1}), the NC one ({\tt iccnc=2}) or the
total process including CC+NC+interference ({\tt  iccnc=3}).
\vsk
{\tt READ*,ips\_cc      }\parno
{\tt READ*,ips\_nc      }\parno
{\tt ips\_cc} and {\tt ips\_nc} allow to choose among the various phase space
mappings for the integration. {\tt ips\_cc} refers to the phase space
of CC or Higgs signal contributions. {\tt ips\_nc} to NC 
contributions. Of course when only CC processes or only Higgs signal are 
considered {\tt ips\_nc} is irrelevant. The
same happens to {\tt ips\_cc} when only neutral processes without Higgs are 
computed.
Both  parameters  can assume 3 values: {\tt 1} for double resonant mapping, 
{\tt 2} for single resonant, {\tt 3} for non resonant. When {\tt 2} 
is chosen the invariant mass
over which a transformation is performed to take care of the resonant peak is
that formed by particles 3 and 4. 
\vsk
{\tt READ*,icc3        }\parno
 yes/no CC3 contribution only. When {\tt icc3} is set to {\tt 1} only the 
   three double resonant diagrams (CC3) corresponding to $WW$ production and 
   decay are computed. If {\tt icc3=0} all CC contributions are computed.
\vsk
{\tt READ*,isr         }\parno
 yes/no ISR:
initial state radiation (ISR) is included when {\tt isr=1}, not computed if 
{\tt isr=0}.
\vsk
{\tt READ*,ipr         }\parno
 yes/no running widths: 
   this flag selects among running or constant $Z$, $W$, 
Higgs widths in s-channel
    propagators:\par
          {\tt ipr=0}  $Z$, $W$, Higgs boson constant width\par
          {\tt ipr=1}  $Z$, $W$, Higgs boson s-dependent width
\vsk
{\tt READ*,iswgcomp    }\parno
 yes/no $sin^2\theta_W$ and $g$ computed. 
   If this flag is set to {\tt 1}, it is used the renormalization scheme in 
   which $sin^2\theta_W$ and $g$ are computed from $Z$ mass, $W$ mass, $G_f$.
   If it is set to {\tt 0}, the values for $sin^2\theta_W$ and 
   $\alpha_{em}$ are taken from the {\tt DATA}. The relation between $g$ and
   $\alpha_{em}$  
   is always  $g^2=4 \pi \alpha_{em} / sin^2 \theta_W$.
\vsk
{\tt READ*,igwcomp,igzcomp,ighcomp}\parno
 yes/no $W$, $Z$, H width computed.
   When one of these flags is {\tt =1}, the corresponding $W$, $Z$ or Higgs 
   width
   is computed by standard formulas. If it is {\tt =0}, the value for the 
   corresponding width is the one given in the {\tt DATA}.
\vsk
{\tt READ*,icoul       }\parno
   Coulomb corrections may ({\tt icoul=1}) or not ({\tt icoul=0}) be computed.
\vsk
{\tt READ*,istrcor     }\parno
 yes/no 'naive' QCD corrections to the cross sections. It has to be noticed
 that, if {\tt istrcor=0} and quarks are present in the final state, 
the eventual  width computations 
 (performed when  {\tt igwcomp,  igzcomp or ighcomp} is {\tt =1}) do not 
 include QCD corrections. 
\vsk
{\tt READ*,iqu         }\parno
 yes/no QCD diagrams for 4-quarks NC. The order at which one computes four
 fermion diagrams is $ \alpha ^2$. There are some diagrams in NC processes
 $O(\alpha  \alpha_s )$ whose contribution can be important and can be
 included choosing {\tt iqu=1}. 
\vsk
{\tt READ*,icut        }\parno
{\tt IF(icut.EQ.1)THEN}\parno
{\tt \hsk  READ*,e\_min     }\parno
{\tt \hsk  READ*,e\_max     }\parno
{\tt \hsk  READ*,rm\_min    }\parno
{\tt \hsk  READ*,rm\_max    }\parno
{\tt \hsk  READ*,pt\_min    }\parno
{\tt \hsk  READ*,pt\_max    }\parno
{\tt \hsk  READ*,icos      }\parno
{\tt \hsk  READ*,thbeam\_min}\parno
{\tt \hsk  READ*,thbeam\_max}\parno
{\tt \hsk  READ*,thsep\_min }\parno
{\tt \hsk  READ*,thsep\_max }\parno
{\tt ENDIF}\parno
   Cuts may ({\tt icut=1}) or not ({\tt icut=0}) be implemented. If 
   {\tt icut=1} all default cuts of the above list must be specified. \parno
   {\tt e\_min} and {\tt e\_max} correspond to the 4 lower and upper
   energies for particle 3, 4, 5, 6 respectively. \parno
   {\tt rm\_min}  and {\tt rm\_max} are the  6 invariant mass lower and upper
    limits respectively. They must be given in the following order:
    m(34), m(35), m(36), m(45), m(46), m(56).\parno
   {\tt pt\_min}  and {\tt pt\_max} are the 4 lower and upper values of the
   transverse momenta. The order is as before 3, 4, 5, 6.\parno
   {\tt icos = 1} implies that the following angular cuts must be expressed in
    terms of the cosines of the angles. With {\tt icos = 0} one must instead
    specify the angles in degrees.\parno
   {\tt thbeam\_min} and {\tt thbeam\_max} are the 4 lower and the 4 upper
   limits for the angle that particles 3,4,5,6 produce with the beam ($e^+$). 
   \parno
  {\tt thsep\_min}  and {\tt thsep\_max} are  the 6 lower and the 6 upper
   limits for  particle-particle angles: the order is again (3 4), (3 5),
   (3 6), (4 5), (4 6), (5 6).
\vsk
{\tt READ*,ianc        }\parno
{\tt IF(ianc.EQ.1)THEN}\parno
{\tt  \hsk READ*, delz,xf,xz,yf,yz,zz }\parno
{\tt ENDIF}\parno
   Anomalous Couplings contributions may ({\tt ianc=1}) or not ({\tt ianc=0}) 
   be computed.
   If one wants to compute them, he must also specify the values of the
   parameters $\delta_Z, x_\gamma, x_Z, y_\gamma, y_Z, z_Z$ defined in 
   ref.~\cite{ancoup}
\vsk
{\tt READ*,idistr      }\parno
{\tt IF(idistr.EQ.1)THEN}\parno
{\tt \hsk  READ*,ndistr    }\parno
{\tt \hsk  DO i=1,ndistr}\parno
{\tt \hsk\hsk    READ*,nsubint(i)  }\parno
{\tt \hsk\hsk    READ*,(distr\_estrinf(i,j),j=1,nsubint(i)+1) }\parno
{\tt  \hsk\hsk   READ*,(nbin\_number(i,j),j=1,nsubint(i)) }\parno
{\tt \hsk  ENDDO }\parno
{\tt ENDIF}\parno
Distributions at  parton level can be easily implemented.
If one wants to use this possibility the flag {\tt idistr} must be {\tt =1},
{\tt =0} otherwise.
In the program there is the line:\parno
  {\tt * \hsk\hsk   include 'abdis.dis'}\parno
  which must be uncommented before compiling if one choose {\tt idistr=1}.
  In this case  the file {\tt abdis.dis}
  must be written for implementing distributions, as in the following
   example:
\begin{verbatim}
        string(1)='Distribution: bb~ invariant mass'
        distr_var(1)=sqrt((p3(0)+p4(0))**2-(p3(1)+p4(1))**2-
     &  (p3(2)+p4(2))**2-(p3(3)+p4(3))**2)
        string(2)='Distribution: Charged lepton energy '
        distr_var(2)=p5(0)
\end{verbatim}
   In it, the title of the ith distribution is given in the {\tt character*60
   string(i)}  and the ith quantity to be distributed in bins, computed from
   the particle momenta, is assigned to {\tt distr\_var(i)}.
   The resulting cross sections  corresponding to every single bin will be
   stored in the file {\tt abdis.dat}. Each line will contain 3 numbers:
   the value of  the central point of the bin, the distribution  for the bin
   (cross section divided by the width of the bin, in order to reproduce 
   $d \sigma / dx$ for a distribution of the variable $x$) and the 
   estimated statistical error.\parno
{\tt ndistr} is the number of distributions defined in {\tt abdis.dis}.
\parno For each distribution {\tt i},  one must specify:\par
{\tt nsubint(i)},  the
   number of sub-intervals with different binning in the ith distribution
   (=1 when all bins are of the same length). The subintervals must be
   contiguous.\par
{\tt  (distr\_estrinf(i,j),j=1,nsubint(i)+1)}, the
   lower limits of each subinterval (which coincide with the upper limit of the
   previous one as they must be contiguous) and, as last entry, the upper
   limit of the last subinterval. In case all bins are of the same length,
   this corresponds only to the lower and upper limit of the interval for the
   distribution.\par
{\tt  (nbin\_number(i,j),j=1,nsubint(i))},  the number of bins in each
subinterval.
\vsk
{\tt READ*,iflat        }\parno
{\tt IF(iflat.EQ.1)THEN}\parno
{\tt  \hsk READ*,scalemax   }\parno
{\tt  \hsk READ*,istorvegas }\parno
{\tt \hsk  READ*,irepeat    }\parno
{\tt \hsk  IF(irepeat.eq.2)THEN}\parno
{\tt \hsk\hsk    READ*, nflevts }\parno
{\tt \hsk  END IF}\parno
{\tt \hsk  READ*,istormom   }\parno
{\tt \hsk  READ*,ijetset    }\parno
{\tt \hsk  IF(iproc.GE.6.AND.iproc.LE.8)THEN}\parno
{\tt \hsk\hsk    READ*,interf   }\parno
{\tt \hsk  ENDIF}\parno
{\tt ENDIF}\parno
One may choose ({\tt iflat=1}) or not ({\tt iflat =0}) to generate unweighted
events. In the first case, the number of iterations ({\tt itmx}) which must
be specified in the following  must be {\tt 2}. The integration routine will 
perform the requested number of iterations for thermalization 
(see {\tt iterm, ncall\_term,itmx\_term} below) and
  then the two iterations in which the integral is evaluated.
  In the first iteration the maximum for the hit-or-miss procedure will be
  determined and used in the second iteration where the unweighted generation 
  will  take place. After the run the .log file will report as usual the result
  of the integration and its error. It will also report the maximum used for
  the hit-or-miss procedure, the maximum found in the second iteration,
  and the number of events which were greater than the maximum used.\parno
  There is also the possibility to repeat the generation just starting directly
  from the second iteration. This might be useful if too many events exceeded
  the maximum chosen, or to generate a predetermined number
  of events. 
\vsk
{\tt  scalemax} is the coefficient by which 
  the maximum of the first iteration can be multiplied, in order to vary the 
  efficiency of the hit-or-miss procedure
  or in order to avoid values exceeding the maximum. \parno
  {\tt VEGAS} data  are (if {\tt istorvegas=1}) or not (if {\tt istorvegas=0})
  stored after the first iteration  in {\tt ABVEGAS.DAT} 
  (in {\tt ABVEGAS\_CC.DAT}
  and {\tt ABVEGAS\_NC.DAT} for Mixed and Higgs+background processes). 
  Stored {\tt VEGAS} data are necessary if one 
  wants to rerun the program
  to generate again unweighted events. When the program is
  rerun using {\tt VEGAS} data stored, the maximum of the second
  iteration will be automatically used as the new maximum.\parno
  {\tt irepeat} has to be set to {\tt 0} for the first run.
  It has to be set to {\tt 1} if one wants to rerun exactly with the same
  input starting from the second iteration. In this case the
  same weighted points will be reproduced.
  {\tt irepeat=2}  has to be chosen if one wants  to rerun with the same input
   and grid as  before, but letting the program run until a requested number of
  events {\tt nfltevts} is reached. For both cases {\tt irepeat=1 and 2}
  one might of course vary   {\tt scalemax}, {\tt ijetset} and {\tt istormom}
  with respect to the first run with {\tt irepeat=0}.
  \parno
  The momenta of the unweighted  events are written in {\tt ABMOM.DAT} (or 
  {\tt ABMOM\_SIGN.DAT} and  {\tt ABMOM\_BACK.DAT})  file if 
  {\tt istormom=1}, they are not 
  written if {\tt istormom=0}.  {\tt ABMOM\_SIGN.DAT} is used for CC or 
  Higgs events, {\tt ABMOM\_BACK.DAT} for NC events when mixed or 
   Higgs+background processes are 
  computed.\parno
  Every unweighted event is passed to the standard {\tt COMMON HEPEVT}.
  If {\tt ijetset=1} the subroutine {\tt AB\_LU4FRM} is called for interfacing
  Jetset.  If {\tt ijetset=0}, {\tt HEPEVT} is still filled but the 
  interface to Jetset not called.\parno
  One may choose to produce events for mixed processes in which the
  interference is added to CC ({\tt interf= 1}), or to NC ({\tt interf=0}).
  This implies that also the interference will be considered to have the same
  colour structure of CC or NC respectively (see section 5.3 of 
  ref.~\cite{tb}. 
\vsk
{\tt READ*,acc         }\parno
{\tt READ*,iterm       }\parno
{\tt READ*,ncall\_term  }\parno
{\tt READ*,itmx\_term   }\parno
{\tt READ*,ncall       }\parno
{\tt READ*,itmx        }\parno
These parameters specify how the 
integration will be performed by {\tt VEGAS}.\parno
{\tt acc} is the
   integration accuracy. When this accuracy is reached after a certain
   integration iteration, the remaining iterations are not performed.\parno
   If {\tt iterm=1} a certain number of integration  iterations 
   ({\tt =itmx\_term)} are used only for adapting the integration grid. 
   Their result is not  used for the final integral. Each thermalizing 
   iteration makes use of a maximum of {\tt ncall\_term} evaluations of the 
   amplitude.   If {\tt iterm=0} these iterations are skipped.
   One or two {\tt itmx\_term} with few {\tt ncall\_term} are often useful.
   \parno
   {\tt itmx} is the maximum number of iterations used to evaluate the integral.
   Each iteration makes use of a maximum of {\tt ncall} evaluations of the
   amplitude. Normally it is better not to use more than about five iterations.
   If higher precision is requested it is convenient to increase {\tt ncall} 
   and not   {\tt itmx}\parno
   As a final remark about the choice of these parameters, one must be aware
    of the fact that final results with a $\chi^2$ much greater than the 
   number of  iterations  are not to be trusted. When this happens, one has to
   increase {\tt ncall}.

\vskip 1cm
   In addition to the parameters  in input, other parameters are fixed in the
main program by the following {\tt DATA} statement and may be eventually
changed.
\begin{verbatim}
      DATA rmw/80.26d0/, rmz/91.1884d0/, rmt/175.d0/, rmc/0.75d0/,
     &     rmtau/1.78d0/, rmb_run/2.9d0/,
     &     gamw/2.08d0/, gamz/2.4974d0/, gamh/1.278d-03/,
     &     gf/1.1663892199930875d-05/, alfainv/128.07d0/,
     &     alfas_cc/0.1255d0/, alfas_nc/0.1230d0/,
     &     s2w/0.231030912451068d0/, rms/1000.d0/
\end{verbatim}
\vsk
{\tt rmw, rmz, rmt, rmc, rmtau } are respectively the  $W$,  $Z$, top,
c and  tau masses.
\parno
{\tt rmb\_run } is the quark $b$  mass used for the Higgs coupling.
\parno
{\tt gamw}, {\tt gamz}, {\tt gamh} are the total $W$, $Z$ and Higgs width.
\parno
{\tt gf } is the Fermi coupling constant;
{\tt alfainv } is $1/\alpha_{em}$ at the appropriate scale; {\tt alfas\_cc }
and {\tt alfas\_nc } are $\alpha_s(M_W)$ and $\alpha_s(M_Z)$.
\parno
{\tt s2w } is the Weinberg $\sin^2(\theta_W)$ and 
{\tt rms }  the SUSY scale.

\subsection{Program implementation}
\noindent
In this section, \wph subroutines and functions are briefly described.
\begin{quote}{\footnotesize \begin{verbatim}
      double precision FUNCTION fxn(x,wgt)
      COMMON/abpara/rmx1,gamx1,gx1,rmx2,gamx2,gx2,beta,rlim,s_col,
     &  x1_min,x2_min,xm,smin,emcoupl,estrinf1,estrinf2,estrmed2,
     &  rmx3,gamx3,gx3
      COMMON/abparb/rmw2,gamw,rmz2,rmt2,rcotw,pi,alfa_me,qcdcoupl,
     &            qcdcor_cc,qcdcor_nc
      COMMON/abopzi/isr,ipr,ips,iccnc,iproc,ich,ichcj,ips_cc,ips_nc,
     &  icos,icch,isusy,icut,igwcomp,igzcomp,ighcomp,iswgcomp
      COMMON/abcoup/fer,fel,zer,zel,f3l,f4l,f5l,f6l,z3l,z4l,z5l,z6l,
     &  f3r,f4r,f5r,f6r,z3r,z4r,z5r,z6r,wcl,delz,xf,xz,yf,yz,zz
      COMMON/abflag/icc,icc3,i3e,i4e,i3q,i5q,iqu,i34e,i56ve,ibbveve,iid,
     &  imix,icoul,istrcor,idownl,idownr,ianc
      COMMON/abhigg/rmb,rmb2,rmh,rmh2,gamh,rhzz,rhww,rhbb
      COMMON/absusy/rma,rma2,rzha
      COMMON/abcuts/e_min(3:6),e_max(3:6),thbeam_min(3:6),
     &  thbeam_max(3:6),thsep_min(6),thsep_max(6),rm_min(6),rm_max(6),
     &  beamcut_min(3:6),beamcut_max(3:6),sepcut_min(6),sepcut_max(6),
     &  rm_min2(6),rm_max2(6),pt_min(3:6),pt_max(3:6),e_cm
      COMMON/abstat/ncall_eff
      COMMON/abdist/distr_estrinf(ndismax,nestrmax),bin_width(ndismax,
     &  nintmax),distr_local(ndismax,nbinmax,nitmax),distr_loc_mix
     &  (ndismax,nbinmax,nitmax),dev_local(ndismax,nbinmax,nitmax),
     &  dev_loc_mix(ndismax,nbinmax,nitmax),tail_local(ndismax,nitmax),
     &  tail_loc_mix(ndismax,nitmax)
      COMMON/abidis/ncallbin(ndismax,nbinmax,nitmax),
     &  ncallbin_mix(ndismax,nbinmax,nitmax),nbin_number(ndismax,
     &  nintmax),nbin_sum(ndismax,nintmax),nsubint(ndismax),ndistr,
     &  idistr,it1,it2,init
      COMMON/abcdis/string(ndismax)
      COMMON/abflat/rmaxfxn,rmaxfxn_1it,rmaxfxn_2it,rmaxfxn_cc_1it,
     &  rmaxfxn_cc_2it,rmaxfxn_nc_1it,rmaxfxn_nc_2it,scalemax
      COMMON/abifla/itmx,novermax,iflat,iseed,istorvegas,istormom,iterm,
     &  ijetset,interf
      COMMON/abfla2/irepeat,nevent,nflevts
      COMMON /abrann/ idum
      COMMON/HEPEVT/NEVHEP,NHEP,ISTHEP(NMXHEP),IDHEP(NMXHEP),
      COMMON/HEPEVT/NEVHEP,NHEP,ISTHEP(NMXHEP),IDHEP(NMXHEP),
     &  JMOHEP(2,NMXHEP),JDAHEP(2,NMXHEP),PHEP(5,NMXHEP),VHEP(4,NMXHEP)
      EXTERNAL ee_4f,ee_bbvv,ee_bbmumu,ee_bbbb,ee_bbee
\end{verbatim}}\end{quote}
\noindent
The function {\tt fxn} is called by {\tt VEGAS} and evaluates the phase space
and ISR.
After momenta reconstruction, it eventually implements the cuts and then calls
the appropriate functions to compute the matrix elements. It also performs
all calculations relative to distributions, it fills the common {\tt HEPEVT},
and if necessary it calls the interface to Jetset, {\tt AB\_LU4FRM}.
\begin{quote}{\footnotesize \begin{verbatim}

      double precision FUNCTION ee_4f(p1,p2,p3,p4,p5,p6)
      COMMON/abparb/rmw2,gamw,rmz2,rmt2,rcotw,pi,alfa_me,qcdcoupl,
     &  qcdcor_cc,qcdcor_nc
      COMMON/abparc/czipr,ccz,cwipr,ccw,chipr,cch,caipr,cca
      COMMON/abcoup/fer,fel,zer,zel,f3l,f4l,f5l,f6l,z3l,z4l,z5l,z6l,
     &       f3r,f4r,f5r,f6r,z3r,z4r,z5r,z6r,wcl,delz,xf,xz,yf,yz,zz
      COMMON/abflag/icc,icc3,i3e,i4e,i3q,i5q,iqu,i34e,i56ve,ibbveve,iid,
     &  imix,icoul,istrcor,idownl,idownr,ianc
\end{verbatim}}\end{quote} 
\noindent
{\tt ee\_4f} computes the amplitude for all the processes with massless four
fermions in the final state. It also evaluates Coulomb, QCD corrections 
and Anomalous Couplings. It is called by {\tt fxn}.
\begin{quote}{\footnotesize \begin{verbatim}

      double precision FUNCTION ee_bbvv(p1,p2,p3,p4,p5,p6)
      COMMON/abhigg/rmb,rmb2,rmh,rmh2,gamh,rhzz,rhww,rhbb
      COMMON/abparb/rmw2,gamw,rmz2,rmt2,rcotw,pi,alfa_me,qcdcoupl,
     &  qcdcor_cc,qcdcor_nc
      COMMON/abparc/czipr,ccz,cwipr,ccw,chipr,cch,caipr,cca
      COMMON/abcoup/fer,fel,zer,zel,f3l,f4l,f5l,f6l,z3l,z4l,z5l,z6l,
     &  f3r,f4r,f5r,f6r,z3r,z4r,z5r,z6r,wcl,delz,xf,xz,yf,yz,zz
      COMMON/abcopl/zvl,zvr,fqdl,fqdr,zqdl,zqdr,fqul,fqur,zqul,zqur
      COMMON/abflag/icc,icc3,i3e,i4e,i3q,i5q,iqu,i34e,i56ve,ibbveve,iid,
     &  imix,icoul,istrcor,idownl,idownr,ianc
\end{verbatim}}\end{quote} 
\noindent
{\tt ee\_bbvv} is called by {\tt fxn} and gives the massive amplitude 
relative to the processes:

$e^+e^-\ar b\bar b \nu_l\bar\nu_l$ \quad \quad $l$=($e,\mu,\tau$).
\begin{quote}{\footnotesize \begin{verbatim}

      double precision FUNCTION ee_bbmumu(p1,p2,p3,p4,p5,p6)
      COMMON/abhigg/rmb,rmb2,rmh,rmh2,gamh,rhzz,rhww,rhbb
      COMMON/abparb/rmw2,gamw,rmz2,rmt2,rcotw,pi,alfa_me,qcdcoupl,
     &  qcdcor_cc,qcdcor_nc
      COMMON/abparc/czipr,ccz,cwipr,ccw,chipr,cch,caipr,cca
      COMMON/abcoup/fer,fel,zer,zel,f3l,f4l,f5l,f6l,z3l,z4l,z5l,z6l,
     &  f3r,f4r,f5r,f6r,z3r,z4r,z5r,z6r,wcl,delz,xf,xz,yf,yz,zz
      COMMON/abcopl/zvl,zvr,fqdl,fqdr,zqdl,zqdr,fqul,fqur,zqul,zqur
      COMMON/abflag/icc,icc3,i3e,i4e,i3q,i5q,iqu,i34e,i56ve,ibbveve,iid,
     &  imix,icoul,istrcor,idownl,idownr,ianc
\end{verbatim}}\end{quote} 
\noindent
{\tt ee\_bbmumu} is called by {\tt fxn} and computes the massive amplitude 
relative to the processes:

$e^+e^-\ar b\bar b q\bar q$ \quad {$q$=($u$,$d$,$c$,$s$)}

$e^+e^-\rightarrow b\bar b l^-l^+$  \quad {$l$=($\mu,\tau$)}
\begin{quote}{\footnotesize \begin{verbatim}

      double precision function ee_bbee(q1,q2,q3,q4,q5,q6)
      COMMON/abhigg/rmb,rmb2,rmh,rmh2,gamh,rhzz,rhww,rhbb
      COMMON/abparb/rmw2,gamw,rmz2,rmt2,rcotw,pi,alfa_me,qcdcoupl,
     &  qcdcor_cc,qcdcor_nc
      COMMON/abparc/czipr,ccz,cwipr,ccw,chipr,cch,caipr,cca
      COMMON/abcoup/fer,fel,zer,zel,f3l,f4l,f5l,f6l,z3l,z4l,z5l,z6l,
     &  f3r,f4r,f5r,f6r,z3r,z4r,z5r,z6r,wcl,delz,xf,xz,yf,yz,zz
      COMMON/abcopl/zvl,zvr,fqdl,fqdr,zqdl,zqdr,fqul,fqur,zqul,zqur
      COMMON/abflag/icc,icc3,i3e,i4e,i3q,i5q,iqu,i34e,i56ve,ibbveve,iid,
     &  imix,icoul,istrcor,idownl,idownr,ianc
\end{verbatim}}\end{quote} 
\noindent
{\tt ee\_bbee} is called by {\tt fxn} and gives the massive amplitude 
relative to the process :

$e^+e^-\rightarrow b\bar b e^-e^+$  
\begin{quote}{\footnotesize \begin{verbatim}

      double precision FUNCTION ee_bbbb(p1,p2,p3,p4,p5,p6)
      COMMON/abopzi/isr,ipr,ips,iccnc,iproc,ich,ichcj,ips_cc,ips_nc,
     &  icos,icch,isusy,icut,igwcomp,igzcomp,ighcomp,iswgcomp
      COMMON/abparb/rmw2,gamw,rmz2,rmt2,rcotw,pi,alfa_me,qcdcoupl,
     &  qcdcor_cc,qcdcor_nc
      COMMON/abparc/czipr,ccz,cwipr,ccw,chipr,cch,caipr,cca
      COMMON/abcoup/fer,fel,zer,zel,f3l,f4l,f5l,f6l,z3l,z4l,z5l,z6l,
     &  f3r,f4r,f5r,f6r,z3r,z4r,z5r,z6r,wcl,delz,xf,xz,yf,yz,zz
      COMMON/abcopl/zvl,zvr,fqdl,fqdr,zqdl,zqdr,fqul,fqur,zqul,zqur
      COMMON/abflag/icc,icc3,i3e,i4e,i3q,i5q,iqu,i34e,i56ve,ibbveve,iid,
     &  imix,icoul,istrcor,idownl,idownr,ianc
      COMMON/abhigg/rmb,rmb2,rmh,rmh2,gamh,rhzz,rhww,rhbb
      COMMON/absusy/rma,rma2,rzha

\end{verbatim}}\end{quote} 
\noindent
{\tt ee\_bbbb} computes the massive amplitude relative to the process 
$e^+e^-\rightarrow b\bar b b \bar b$ .
It contains both SM ($hZ$) and MSSM ($hZ$ + $hA$) Higgs diagrams.
It is called by {\tt fxn}.
\begin{quote}{\footnotesize \begin{verbatim}

      Subroutine AB_LU4FRM(ichar,IRAD,ITAU,IERR)
      COMMON/LUJETS/N,K(4000,5),P(4000,5),V(4000,5)
      COMMON/LUDAT1/MSTU(200),PARU(200),MSTJ(200),PARJ(200)
\end{verbatim}}\end{quote} 
\noindent
This routine is an interface to Jetset that we have produced modifying the
subroutine {\tt LU4FRM} by T. Sj\"ostrand to adapt it to {\tt WPHACT}.
The comment lines of the routine {\tt AB\_LU4FRM} may be useful.
\begin{quote}{\footnotesize \begin{verbatim}

      SUBROUTINE vegas(region,ndim,fxn,init,ncall,itmx,nprn,tgral,sd,
     *  chi2a,acc,xi,it,ndo,si,swgt,schi)
      EXTERNAL fxn
      COMMON/abresl/resl(10),standdevl(10)
      COMMON /abrann/ idum
      COMMON/abchia/calls
      COMMON/abstat/ncall_eff
      COMMON/abfla2/irepeat,nevent,nflevts
\end{verbatim}}\end{quote} 
\noindent
This subroutine by P. Lepage \cite{vegas} performs the 
multidimensional integrations.
It has been modified in order to obtain unweighted event generation, 
distributions and separated integration of CC ( or Higgs) and NC contributions
both for MIX and massive $b$ processes. It is called by the main program and
it calls the function {\tt fxn}.
\begin{quote}{\footnotesize \begin{verbatim}

      FUNCTION ran2(idum)
      COMMON/absalv/iv,iy,idum2
\end{verbatim}}\end{quote} 
\noindent	
This function is a random number generator. Both {\tt VEGAS} and \wph make
use of it.
\begin{quote}{\footnotesize \begin{verbatim}

      SUBROUTINE rebin(rc,nd,r,xin,xi)
\end{verbatim}}\end{quote} 
\noindent
This routine is used by {\tt VEGAS}.
\begin{quote}{\footnotesize \begin{verbatim}

      double precision FUNCTION gammln(xx)
\end{verbatim}}\end{quote} 
\noindent
It computes the logarithm of the $\Gamma$ function entering in the electron
structure functions.
\vsk
\begin{quote}{\footnotesize \begin{verbatim}

      SUBROUTINE initialize
      COMMON/abcoup/fer,fel,zer,zel,f3l,f4l,f5l,f6l,z3l,z4l,z5l,z6l,
     &  f3r,f4r,f5r,f6r,z3r,z4r,z5r,z6r,wcl,delz,xf,xz,yf,yz,zz
      COMMON/abcopl/zvl,zvr,fqdl,fqdr,zqdl,zqdr,fqul,fqur,zqul,zqur
      COMMON/abflag/icc,icc3,i3e,i4e,i3q,i5q,iqu,i34e,i56ve,ibbveve,iid,
     &  imix,icoul,istrcor,idownl,idownr,ianc
      COMMON/abflat/rmaxfxn,rmaxfxn_1it,rmaxfxn_2it,rmaxfxn_cc_1it,
     &  rmaxfxn_cc_2it,rmaxfxn_nc_1it,rmaxfxn_nc_2it,scalemax
      COMMON/abifla/itmx,novermax,iflat,iseed,istorvegas,istormom,iterm,
     &  ijetset,interf
      COMMON/abfla2/irepeat,nevent,nflevts
      COMMON/abopzi/isr,ipr,ips,iccnc,iproc,ich,ichcj,ips_cc,ips_nc,
     &  icos,icch,isusy,icut,igwcomp,igzcomp,ighcomp,iswgcomp
      COMMON/HEPEVT/NEVHEP,NHEP,ISTHEP(NMXHEP),IDHEP(NMXHEP),
     &  JMOHEP(2,NMXHEP),JDAHEP(2,NMXHEP),PHEP(5,NMXHEP),VHEP(4,NMXHEP)
\end{verbatim}}\end{quote} 
\noindent
For a process chosen by means of input parameters, this subroutine, called
by the main program, computes
all chiral coupling constants and it initializes all  variables 
to be passed to other parts of the program and to the common {\tt HEPEVT}.
\begin{quote}{\footnotesize \begin{verbatim}

      SUBROUTINE printer(rmh,gamh,rma,gama,tgb,rmb)
      COMMON/abopzi/isr,ipr,ips,iccnc,iproc,ich,ichcj,ips_cc,ips_nc,
     &  icos,icch,isusy,icut,igwcomp,igzcomp,ighcomp,iswgcomp
      COMMON/abparb/rmw2,gamw,rmz2,rmt2,rcotw,pi,alfa_me,qcdcoupl,
     &            qcdcor_cc,qcdcor_nc
      COMMON/abcoup/fer,fel,zer,zel,f3l,f4l,f5l,f6l,z3l,z4l,z5l,z6l,
     &  f3r,f4r,f5r,f6r,z3r,z4r,z5r,z6r,wcl,delz,xf,xz,yf,yz,zz
      COMMON/abflag/icc,icc3,i3e,i4e,i3q,i5q,iqu,i34e,i56ve,ibbveve,iid,
     &  imix,icoul,istrcor,idownl,idownr,ianc
      COMMON/abcuts/e_min(3:6),e_max(3:6),thbeam_min(3:6),
     &  thbeam_max(3:6),thsep_min(6),thsep_max(6),rm_min(6),rm_max(6),
     &  beamcut_min(3:6),beamcut_max(3:6),sepcut_min(6),sepcut_max(6),
     &  rm_min2(6),rm_max2(6),pt_min(3:6),pt_max(3:6),e_cm
      COMMON/abinpu/rmw,rmz,rmb_run,rmc,rmc_run,rmtau,gamz,gf,s2w,
     &  alfainv,alfas_cc,alfas_nc
      COMMON/abflat/rmaxfxn,rmaxfxn_1it,rmaxfxn_2it,rmaxfxn_cc_1it,
     &  rmaxfxn_cc_2it,rmaxfxn_nc_1it,rmaxfxn_nc_2it,scalemax
      COMMON/abifla/itmx,novermax,iflat,iseed,istorvegas,istormom,iterm,
     &  ijetset,interf
      COMMON/abfla2/irepeat,nevent,nflevts
\end{verbatim}}\end{quote} 
\noindent
This subroutine is devoted to write in the output file all the essential
informations about the selected process, input, options and variables defined
in the {\tt DATA}. It is called by the main program.

\section{Conclusions}
We have described version 1.0 of \wph. The program computes all processes
with a 4 fermion final state. All its main features  and options are 
listed in the Program Summary. The way of using it is mainly explained in
the input section \ref{input}, and some useful examples can be found in
the test runs.

\wph is a MC integrator and event generator portable and self contained. 
For the amplitudes computation it  makes use of a rather
new helicity method which allows fast evaluations. This we consider to be
an  essential ingredient, together with careful mappings of the integration
variables, in order to combine the usual advantages of MC's 
with high precision and  reliability.

\section*{Acknowledgments}
We are extremely grateful to Dima Bardin and Giampiero Passarino.
With them we had continuous discussions and comparisons
among  our program, {\tt GENTLE/4fan}  and {\tt WTO}. The intrinsic
difference of the three approaches has allowed the most useful
tests. In particular {\tt WTO} and \wph had the unique chance to be developed
contemporarily and we have surely benefited of the intense day by day 
collaboration with Giampiero Passarino.
We like to thank all our colleagues of the LEP2 WW working group and in
particular the two coordinators Ronald Kleiss and Dima Bardin
for the  live atmosphere of the group as well as for comparisons and 
exchange of informations. We gratefully acknowledge in particular friendly 
and useful discussions with Roberto Pittau.  Finally we thank Roberto
Contri for his  advises and collaboration in testing {\tt WPHACT}.

\vfill\eject

\section*{Test runs }

We report here two significant test runs for CC and NC+Higgs   
processes with different options.
\vsk
\leftline{\bf{Run 1:}}
\noindent
This run computes the cross section and two distributions for the full CC20 
process
$e^+ e^-\rightarrow e^-\bar\nu_e u \bar d$. A double resonant phase space 
on the $W$ masses has been chosen to perform the integration over the 9 
variables. ISR, Coulomb and QCD corrections have been
included as well as Anomalous Couplings. The following cuts are implemented:
 $E_{e^-,u,\bar d}\ge 20$ GeV,\quad  $M_{u\bar d}\ge 10$ GeV,\quad
$|cos(\theta_{beam})|_{e^-,u,\bar d}\le 0.9$, 
\quad $|cos(\theta_{u\bar d})|\le 0.9$. \parno
Distributions of the $M_{W^+}$ invariant mass and the outgoing electron 
energy are calculated. The corresponding plots are reported in figs. 1,2.
 They  contain 51 and 150 bins respectively.  The relative 
statistical errors are a few permill for every bin. 
They are not visible in the figures.
\parno
The relative error obtained is $ 0.019 \% $. The total CPU time 
on AlphaStation 600 5/333 is $24$ min and 
$36$ sec, which corresponds to $0.6E-04$ sec. per call.

\vsk
Input file:
\begin{quote}{\footnotesize \begin{verbatim}
$r wphact
190.d0                             ! centre of mass energy
4                                  ! selects the kind of process
1                                  ! selects the channel
1                                  ! CC( or Higgs signal) phase space
1                                  ! NC phase space
0                                  ! yes/no CC3 only
1                                  ! yes/no ISR
1                                  ! yes/no running widths
0                                  ! yes/no s2w and g computed
1 0 0                              ! yes/no W, Z, Higgs width computed
1                                  ! yes/no Coulomb corrections
1                                  ! yes/no QCD corrections
0                                  ! yes/no QCD diagrams
1                                  ! yes/no cuts
20.d0 0.d0 20.d0 20.d0             ! 4 energy lower cuts
300.d0 300.d0 300.d0 300.d0        ! 4 energy upper cuts
0.d0 0.d0 0.d0 0.d0 0.d0 10.d0     ! 6 invariant mass lower limits
300.d0 300.d0 300.d0 300.d0 300.d0 300.d0 !6 invariant mass upper limits
0.d0 0.d0 0.d0 0.d0                ! 4 transverse momenta lower cuts
300.d0 300.d0 300.d0 300.d0        ! 4 transverse momenta upper cuts
1                                  ! angular cuts in deg (0) or cos (1)
0.9d0 1.d0 0.9d0 0.9d0             ! 4 particle-beam angle lower cuts
-0.9d0 -1.d0 -0.9d0 -0.9d0         ! 4 particle-beam angle upper cuts
1.d0 1.d0 1.d0 1.d0 1.d0 0.9d0     ! 6 particle-particle angle lower cuts
-1.d0 -1.d0 -1.d0 -1.d0 -1.d0 -0.9d0 !6 part-part angle upper cuts
1                                  ! yes/no anomalous couplings
-0.5d0 0.d0 0.d0 0.d0 0.d0 0.d0    ! anomalous couplings parameters
1                                  ! yes/no distributions
2                                  ! number of distributions
3                                  ! sub-intervals number with different binning
75.71d0 78.21d0 82.31d0 84.81d0    ! limits of each sub-interval
5 41 5                             ! number of bins in each sub-interval
1                                  ! sub-intervals number with different binning
15.d0 90.d0                        ! limits of each sub-interval
150                                ! number of bins in each sub-interval
0                                  ! yes/no flat generation
0.0001d0                           ! integration accuracy
1                                  ! yes/no thermalization
2000000                            ! thermalization calls per iteration
2                                  ! thermalization iterations 
10000000                           ! integration calls per iteration
2                                  ! integration iterations
$exit
\end{verbatim}}\end{quote}
\vsk
File abdis.dis :

\begin{quote}{\footnotesize \begin{verbatim}
        string(1)='Distribution: Mw+'
        distr_var(1)=sqrt((p5(0)+p6(0))**2-(p5(1)+p6(1))**2-
     &  (p5(2)+p6(2))**2-(p5(3)+p6(3))**2)
        string(2)='Distribution: Lepton energy'
        distr_var(2)=p3(0)
\end{verbatim}}\end{quote}
  
\parno
Output file:

\begin{quote}{\footnotesize \begin{verbatim}
-----------------------------------------------------
 
CC20 )  e-(p3) ve~(p4) u(p5) d~(p6)
WW signal + background
 
INPUT
cm energy             =          0.1900000D+03 GeV
 
DATA
Z mass                =          0.9118840D+02 GeV
W mass                =          0.8026000D+02 GeV
Z width               =          0.2497400D+01 GeV
Gf                    =          0.1166389D-04 GeV-2
s2w                   =          0.2310309D+00
1/alfa_em             =          0.1280700D+03
alfas_cc              =          0.1255000D+00
 
DERIVED QUANTITIES
W width               =          0.2088612D+01 GeV
 
OPTIONS
both Z and W boson s-dependent width
Born + QED
Coulomb corrections included
Naive QCD corrections included
Double resonant "CC" phase space
 
Anomalous coupling parameters:
-----------------------------
 delz = -0.5    xf =  0.0
 xz   =  0.0    yf =  0.0
 yz   =  0.0    zz =  0.0

Cuts :
-----------------
ENERGY_MIN(3,4,5,6)           =(  20.00,  0.00, 20.00, 20.00 ) GeV
ENERGY_MAX(3,4,5,6)           =(  300.00, 300.00, 300.00, 300.00 ) GeV
MASS_MIN(34,35,36,45,46,56)   =(   0.00,  0.00,  0.00,  0.00,  0.00, 10.00 ) GeV
MASS_MAX(34,35,36,45,46,56)   =(  300.00, 300.00, 300.00, 300.00, 300.00, 300.)
PT_MIN(3,4,5,6)               =(   0.00,  0.00,  0.00,  0.00 ) GeV
PT_MAX(3,4,5,6)               =(  300.00, 300.00, 300.00, 300.00 ) GeV
COSBEAM_MAX(3,4,5,6)          =(   0.90,  1.00,  0.90,  0.90 )
COSBEAM_MIN(3,4,5,6)          =(  -0.90, -1.00, -0.90, -0.90 )
COSSEP_MAX(34,35,36,45,46,56) =(   1.00,  1.00,  1.00,  1.00,  1.00,  0.90 )
COSSEP_MIN(34,35,36,45,46,56) =(  -1.00, -1.00, -1.00, -1.00, -1.00, -0.90 )
-----------------------------------------------------
 
 
Thermalization

input parameters for vegas:  ndim=  9  ncall=    1835008.
                             it=    1  itmx=    2

iteration no.  1:            effective ncall=    1052601
iteration no.  1: integral = 0.4101226    +/-  0.50E-03
all iterations:   integral = 0.4101226    +/- 0.503E-03 chi**2/it'n = 0.12E-07

iteration no.  2:            effective ncall=    1273154
iteration no.  2: integral = 0.4100550    +/-  0.29E-03
all iterations:   integral = 0.4100716    +/- 0.249E-03 chi**2/it'n = 0.13E-01
 
CC process

input parameters for vegas:  ndim=  9  ncall=    9765625.
                             it=    1  itmx=    2

iteration no.  1:            effective ncall=    6914065
iteration no.  1: integral = 0.4101164    +/-  0.11E-03
all iterations:   integral = 0.4101164    +/- 0.110E-03 chi**2/it'n = 0.00E+00

iteration no.  2:            effective ncall=    6934162
iteration no.  2: integral = 0.4100655    +/-  0.11E-03
all iterations:   integral = 0.4100908    +/- 0.774E-04 chi**2/it'n = 0.11    
 
-----------------------------------------------------
 
Sigma = 0.4100908D+00 +/- 0.774D-04 (pb)

______________________________________________________________________________
\end{verbatim}}\end{quote}

\vsk
\leftline{\bf{Run 2:}}
\noindent
This output refers to NC25 process $ e^+ e^-\rightarrow b \bar b \mu^-
\mu^+$ with massive $b$'s in the final state. The cross section receives
contributions both from Higgs and Background NC diagrams integrated by means of 
two phase spaces accounting for $ZH$ and $ZZ$ peak structure respectively.
ISR and QCD corrections are included. The cuts are:
$M_{b\bar b}\ge 50$ GeV and $M_Z-25$ GeV $\le M_{\mu\mu}\le M_Z+25$ GeV.
This output shows in particular two possible examples of unweighted events 
generation:  the first run produces unweighted events (which are 
stored in {\tt ABMOM\_ SIGN(BACK).DAT})   with an efficiency of a few percent. 
Making use of the Vegas data saved in
{\tt VEGAS\_ CC(NC).DAT} during the first run, a second run 
allows to generate a prefixed number of 50000  unweighted events in a short 
CPU time.
\parno
The relative error obtained is $ 0.0028 \% $. The total CPU time for the both
runs on AlphaStation 600 5/333 is $4$ h $4$ min and $5$ sec.

\vsk
First input file:
\begin{quote}{\footnotesize \begin{verbatim}
$r wphact
190.d0                             ! centre of mass energy
36                                 ! selects the kind of process
1                                  ! selects the channel
2.9d0                              ! quark "b" mass
3                                  ! icch (=1 Higgs, =2 Backg., =3 Higgs+Backg.)
0                                  ! yes/no SUSY
80.d0                              ! Higgs mass
1                                  ! CC( or Higgs) phase space
1                                  ! NC phase space
0                                  ! yes/no CC3 only
1                                  ! yes/no ISR
1                                  ! yes/no running widths
1                                  ! yes/no s2w and g computed
1 0 1                              ! yes/no W, Z, Higgs width computed
0                                  ! yes/no Coulomb corrections
1                                  ! yes/no QCD corrections
0                                  ! yes/no QCD diagrams
1                                  ! yes/no cuts
0.d0 0.d0 0.d0 0.d0                ! 4 energy lower cuts
300.d0 300.d0 300.d0 300.d0        ! 4 energy upper cuts
50.d0 0.d0 0.d0 0.d0 0.d0 66.1888d0 ! 6 invariant mass lower limits
300.d0 300.d0 300.d0 300.d0 300.d0 116.1888d0 !6 invariant mass upper limits
0.d0 0.d0 0.d0 0.d0                ! 4 transverse momenta lower cuts
300.d0 300.d0 300.d0 300.d0        ! 4 transverse momenta upper cuts
0                                  ! angular cuts in deg (0) or cos (1)
0.d0 0.d0 0.d0 0.d0                ! 4 particle-beam angle lower cuts
180.d0 180.d0 180.d0 180.d0        ! 4 particle-beam angle upper cuts
0.d0 0.d0 0.d0 0.d0 0.d0 0.d0      ! 6 particle-particle angle lower cuts
180.d0 180.d0 180.d0 180.d0 180.d0 180.d0 !6 part-part angle upper cuts
0                                  ! yes/no anomalous couplings
0                                  ! yes/no distributions
1                                  ! yes/no flat generation
1.1d0                              ! scale factor for the maximum
1                                  ! yes/no data VEGAS stored
0                                  ! yes/no second iteration repeated
1                                  ! yes/no flat momenta stored
0                                  ! yes/no Jetset program tied
0.0005d0                           ! integration accuracy
1                                  ! yes/no thermalization
1000000                            ! thermalization calls per iteration
2                                  ! thermalization iterations 
20000000                           ! integration calls per iteration
2                                  ! integration iterations
$exit

\end{verbatim}}\end{quote}
\vsk
First output file:

\begin{quote}{\footnotesize \begin{verbatim}
-----------------------------------------------------
 
NC25 )  b(p3) b~(p4) mu-(p5) mu+(p6)
Higgs signal + background
 
INPUT
cm energy             =          0.1900000D+03 GeV
Higgs mass            =          0.8000000D+02 GeV
b mass                =          0.2900000D+01 GeV
 
DATA
Z mass                =          0.9118840D+02 GeV
W mass                =          0.8026000D+02 GeV
c mass                =          0.7500000D+00 GeV
tau mass              =          0.1780000D+01 GeV
Z width               =          0.2497400D+01 GeV
Gf                    =          0.1166389D-04 GeV-2
alfas_nc              =          0.1230000D+00
 
DERIVED QUANTITIES
W width               =          0.2090172D+01 GeV
Higgs width           =          0.1937374D-02 GeV
s2w                   =          0.2253258D+00
1/alfa_em             =          0.1312146D+03
 
OPTIONS
both Z and H boson s-dependent width
Born + QED
Naive QCD corrections included
Double resonant "Higgs signal" phase space
Double resonant "Higgs background" phase space
Cuts :
-----------------
ENERGY_MIN(3,4,5,6)           =(   0.00,  0.00,  0.00,  0.00 ) GeV
ENERGY_MAX(3,4,5,6)           =(  300.00, 300.00, 300.00, 300.00 ) GeV
MASS_MIN(34,35,36,45,46,56)   =(  50.00,  0.00,  0.00,  0.00,  0.00, 66.19 ) GeV
MASS_MAX(34,35,36,45,46,56)   =(  300.00, 300.00, 300.00, 300.00, 300.00, 116.)
PT_MIN(3,4,5,6)               =(   0.00,  0.00,  0.00,  0.00 ) GeV
PT_MAX(3,4,5,6)               =(  300.00, 300.00, 300.00, 300.00 ) GeV
THBEAM_MIN(3,4,5,6)           =(   0.00,  0.00,  0.00,  0.00 ) deg
THBEAM_MAX(3,4,5,6)           =( 180.00,180.00,180.00,180.00 ) deg
THSEP_MIN(34,35,36,45,46,56)  =(   0.00,  0.00,  0.00,  0.00,  0.00,  0.00 ) deg
THSEP_MAX(34,35,36,45,46,56)  =( 180.00,180.00,180.00,180.00,180.00,180.00 ) deg
-----------------------------------------------------
 
Flat events generation
VEGAS data stored in ABVEGAS_CC(NC).DAT
Maximum scale factor = 0.110D+01
Flat events stored in ABMOM_SIGN.DAT
Flat events stored in ABMOM_BACK.DAT

Thermalization

input parameters for vegas:  ndim=  9  ncall=     786432.
                             it=    1  itmx=    2

iteration no.  1:            effective ncall=     743476
iteration no.  1: integral = 0.1661491E-01+/-  0.75E-05
all iterations:   integral = 0.1661491E-01+/- 0.745E-05 chi**2/it'n = 0.00E+00

input parameters for vegas:  ndim=  9  ncall=     786432.
                             it=    1  itmx=    2

iteration no.  1:            effective ncall=     743476
iteration no.  1: integral = 0.7878380E-02+/-  0.61E-05
all iterations:   integral = 0.7878380E-02+/- 0.608E-05 chi**2/it'n = 0.00E+00

iteration no.  2:            effective ncall=     776518
iteration no.  2: integral = 0.7882914E-02+/-  0.37E-05
all iterations:   integral = 0.7881695E-02+/- 0.315E-05 chi**2/it'n = 0.40    
 
Higgs signal

input parameters for vegas:  ndim=  9  ncall=   19531250.
                             it=    1  itmx=    1

iteration no.  1:            effective ncall=   19292845
iteration no.  1: integral = 0.1661659E-01+/-  0.78E-06
all iterations:   integral = 0.1661659E-01+/- 0.784E-06 chi**2/it'n = 0.00E+00
 
Higgs backg + Higgs sign-backg interference

input parameters for vegas:  ndim=  9  ncall=   19531250.
                             it=    1  itmx=    1

iteration no.  1:            effective ncall=   19345150
iteration no.  1: integral = 0.7879024E-02+/-  0.62E-06
all iterations:   integral = 0.7879024E-02+/- 0.621E-06 chi**2/it'n = 0.00E+00
 
Higgs signal

input parameters for vegas:  ndim=  9  ncall=   19531250.
                             it=    2  itmx=    2

iteration no.  2:            effective ncall=   19317393
iteration no.  2: integral = 0.1661579E-01+/-  0.76E-06
all iterations:   integral = 0.1661618E-01+/- 0.545E-06 chi**2/it'n = 0.54    
 
Higgs backg + Higgs sign-backg interference

input parameters for vegas:  ndim=  9  ncall=   19531250.
                             it=    2  itmx=    2

iteration no.  2:            effective ncall=   19355281
iteration no.  2: integral = 0.7877976E-02+/-  0.60E-06
all iterations:   integral = 0.7878485E-02+/- 0.433E-06 chi**2/it'n =  1.5    
 
-----------------------------------------------------
 
Sigma = 0.2449466D-01 +/- 0.696D-06 (pb)
Informations about flat events generation:
         ----------------------           
Maximum after first VEGAS iteration = 0.309D-07
Maximum after second VEGAS iteration = 0.215D-07
Flat events number =    720755
number of function values over maximum =         0
-----------------------------------------------------
\end{verbatim}}\end{quote}

\vsk
Second input file:\parno
We report only the input changed with respect to the first input.
\begin{quote}{\footnotesize \begin{verbatim}
1.d0                               ! scale factor for the maximum
0                                  ! yes/no data VEGAS stored
2                                  ! yes/no second iteration repeated
50000                              ! number of unweighted events generated
1                                  ! yes/no flat momenta stored
\end{verbatim}}\end{quote}
\vsk
Second output file:\parno
We do not report here {\tt INPUT}, {\tt DATA}, {\tt DERIVED QUANTITIES},
{\tt OPTIONS} and cuts which are the same as in the first output.

\begin{quote}{\footnotesize \begin{verbatim}
______________________________________________________________________________
 
NC25 )  b(p3) b~(p4) mu-(p5) mu+(p6)
Higgs signal + background

........
 
-----------------------------------------------------
 
Flat events generation
Maximum scale factor = 0.100D+01
Flat events stored in ABMOM_SIGN.DAT
Flat events stored in ABMOM_BACK.DAT
 
Higgs signal
                                                                              
Flat events number =     33917
 
Higgs backg + Higgs sign-backg interference

Flat events number =     16083
 
-----------------------------------------------------
 
Sigma = 0.2449466D-01 +/- 0.696D-06 (pb)
Informations about flat events generation:
         ----------------------           
Maximum =   1.330188364059000E-008
Flat events number =     50000
number of function values over maximum =         0
-----------------------------------------------------------------------------

\end{verbatim}}\end{quote}

\vfill\eject

\section*{Table Caption}
\bd
\item [Table 1] DEC AlphaStation 600 5/333 CPU time,
 accuracy and effective calls  (in millions) for some   representative four 
 fermion  processes with ISR.
\item [Table 2] Charged Current and Mixed Charged + Neutral Current 
  four fermion processes. {\tt iproc} and {\tt ich} are the two flags by
  which the appropriate final state is singled out in {\tt WPHACT}.
\item [Table 3] Neutral Current and Neutral Current + Higgs 
  four fermion processes. {\tt iproc} and {\tt ich} are the two flags by
  which the appropriate final state is singled out in {\tt WPHACT}.
\ed

\vfill\eject

\begin{table}[hbt]\centering
\begin{tabular}{|c|l|c|c|c|}
\hline 
\multicolumn{5}{|c|}{} \\
\multicolumn{5}{|c|}{\bf CPU time}\\
\hline
process& final state&  calls(M)& precision & hh:mm:ss\\
\hline
\hline
CC10&$\mu^-$ $\bar\nu_\mu$ $u$ $\bar d$  
 &5.4& 0.0002&00:05:09\\
\hline
CC20&$e^-$ $\bar\nu_e$ $u$ $\bar d$  
 & 5.3 &0.0002&00:06:16\\
\hline
Mix56&$d$ $\bar d$ $u$ $\bar u$  
 & 40 &0.0001&00:52:17\\
\hline
Mix56&$e^-$ $e^+$ $\nu_e$ $\bar\nu_e$  
 & 37 &0.0010&01:31:58\\
\hline
NC48&$e^-$ $e^+$ $u$ $\bar u$  
 & 44 &0.0010&01:28:13\\
\hline
NC64&$u$ $\bar u$ $u$ $\bar u$  
 & 27&0.0008&01:00:32\\
\hline
NC144&$e^-$ $e^+$ $e^-$ $e^+$  
 & 47&0.0010&03:39:10\\
\hline
NC21&$b$ $\bar b$ $\nu_e$ $\bar\nu_e$  
 & 11&0.0001&00:15:24\\
\hline
NC25&$b$ $\bar b$ $\mu^-$ $\mu^+$  
 & 22&0.0001&00:54:08\\
\hline
NC84&$b$ $\bar b$ $b$ $\bar b$  
 &24&0.0001&03:49:10\\
\hline
\multicolumn{5}{c}{} \\
\multicolumn{5}{c}{} \\
\multicolumn{5}{c}{\bf Table 1} \\
\end {tabular}
\end {table}

\vskip	2cm
\bc
\begin{tabular}{|c|c|c l c l|}
\hline 
\multicolumn{6}{|c|}{} \\
\multicolumn{6}{|c|}{\bf CC}\\
\hline
process type&iproc& ich & final state& ich & final state \\
\hline
\hline
CC9&1&1&$\mu^-$ $\bar\nu_\mu$ $\nu_\tau$ $\tau^+$  
 &2&$\mu^+$  $\nu_\mu$  $\bar\nu_\tau$  $\tau^-$\\
\hline
CC18&2&1&$e^-$  $\bar\nu_e$  $\nu_\mu$  $\mu^+$ 
 &3&$e^+$  $\nu_e$  $\bar\nu_\mu$  $\mu^-$ \\
& &2&$e^-$  $\bar\nu_e$  $\nu_\tau$  $\tau^+$ 
 &4&$e^+$  $\nu_e$  $\bar\nu_\tau$  $\tau^-$ \\
\hline
CC10&3&1&$\mu^-$  $\bar\nu_\mu$  $u$  $\bar d$ 
 &5&$\tau^-$  $\bar\nu_\tau$  $u$  $\bar d$ \\
& &2&$\mu^-$  $\bar\nu_\mu$  $c$  $\bar s$ 
 &6&$\tau^-$  $\bar\nu_\tau$  $c$  $\bar s$ \\
& &3&$\mu^+$  $\nu_\mu$  $\bar u$  $d$ 
 &7&$\tau^+$  $\nu_\tau$  $\bar u$  $d$ \\
 & &4&$\mu^+$  $\nu_\mu$  $\bar c$  $s$ 
 &8&$\tau^+$  $\nu_\tau$  $\bar c$  $s$ \\
\hline
CC20&4&1&$e^-$  $\bar\nu_e$  $u$  $\bar d$ 
 &3&$e^+$  $\nu_e$  $\bar u$  $d$ \\
& &2&$e^-$  $\bar\nu_e$  $c$  $\bar s$ 
 &4&$e^+$  $\nu_e$  $\bar c$  $s$ \\
\hline
CC11&5&1&$s$  $\bar c$  $u$  $\bar d$ 
 &2&$\bar s$  $c$  $\bar u$  $d$ \\
\hline
\hline
\multicolumn{6}{|c|}{} \\
\multicolumn{6}{|c|}{\bf MIX}\\
\hline
process type& iproc&ich & final state& ich & final state \\
\hline
\hline
MIX19&6&1&$\mu^-$  $\mu^+$  $\nu_\mu$  $\bar\nu_\mu$ 
 &2&$\tau^-$  $\tau^+$  $\nu_\tau$  $\bar\nu_\tau$ \\
\hline
MIX56&7&1&$e^-$  $e^+$  $\nu_e$  $\bar\nu_e$ 
 &&\\
\hline
MIX43&8&1&$d$  $\bar d$  $u$  $\bar u$ 
 &2&$s$  $\bar s$  $c$  $\bar c$ \\
\hline
\multicolumn{6}{c}{ }\\
\multicolumn{6}{c}{\bf Table 2 }\\
\end {tabular}
\vskip .5cm
\ec
\vfill\eject

\bc
\begin{tabular}{|c|c|c l c l|}
\hline 
\multicolumn{6}{|c|}{} \\
\multicolumn{6}{|c|}{\bf NC}\\
\hline
process type& iproc&ich & final state& ich & final state \\
\hline
\hline
NC6&9&1&$\nu_\mu$  $\bar\nu_\mu$  $\nu_\tau$  $\bar\nu_\tau$ 
&&\\
\hline
NC12&10&1&$\nu_\mu$  $\bar\nu_\mu$  $\nu_e$  $\bar\nu_e$
  &2&$\nu_\tau$  $\bar\nu_\tau$  $\nu_e$  $\bar\nu_e$\\
\hline
NC12&11&1&$\nu_\mu$  $\bar\nu_\mu$  $\nu_\mu$  $\bar\nu_\mu$
  &2&$\nu_\tau$  $\bar\nu_\tau$  $\nu_\tau$  $\bar\nu_\tau$\\
\hline
NC36&12&1&$\nu_e$  $\bar\nu_e$  $\nu_e$  $\bar\nu_e$
&&\\
\hline
NC10&13&1&$u$  $\bar u$  $\nu_\mu$  $\bar\nu_\mu$
  &3&$c$  $\bar c$  $\nu_\mu$  $\bar\nu_\mu$\\
&  &2&$u$  $\bar u$  $\nu_\tau$  $\bar\nu_\tau$
  &4&$c$  $\bar c$  $\nu_\tau$  $\bar\nu_\tau$\\
\hline
NC19&14&1&$u$  $\bar u$  $\nu_e$  $\bar\nu_e$
  &2&$c$  $\bar c$  $\nu_e$  $\bar\nu_e$\\
\hline
NC64&15&1&$u$  $\bar u$  $u$  $\bar u$
  &2&$c$  $\bar c$  $c$  $\bar c$\\
\hline
NC32&16&1&$u$  $\bar u$  $c$  $\bar c$
  &&\\
\hline
NC10&17&1&$\mu^-$  $\mu^+$  $\nu_\tau$  $\bar\nu_\tau$
  &2&$\tau^-$  $\tau^+$  $\nu_\mu$  $\bar\nu_\mu$\\
\hline
NC20&18&1&$e^-$  $e^+$  $\nu_\mu$  $\bar\nu_\mu$
  &2&$e^-$  $e^+$  $\nu_\tau$  $\bar\nu_\tau$\\
\hline
NC19&19&1& $\mu^-$  $\mu^+$  $\nu_e$  $\bar\nu_e$ 
  &2& $\tau^-$  $\tau^+$  $\nu_e$  $\bar\nu_e$ \\
\hline
NC24&20&1& $\mu^-$  $\mu^+$  $u$  $\bar u$ 
  &3& $\tau^-$  $\tau^+$  $u$  $\bar u$ \\
&  &2& $\mu^-$  $\mu^+$  $c$  $\bar c$ 
  &4& $\tau^-$  $\tau^+$  $c$  $\bar c$ \\
\hline
NC48&21&1& $e^-$  $e^+$  $u$  $\bar u$ 
  &2& $e^-$  $e^+$  $c$  $\bar c$ \\
\hline
NC19&22&1& $d$  $\bar d$  $\nu_e$  $\bar\nu_e$ 
  &2& $s$  $\bar s$  $\nu_e$  $\bar\nu_e$ \\
\hline
NC10&23&1& $d$  $\bar d$  $\nu_\mu$  $\bar\nu_\mu$ 
  &3& $d$  $\bar d$  $\nu_\tau$  $\bar\nu_\tau$ \\
&  &2& $s$  $\bar s$  $\nu_\mu$  $\bar\nu_\mu$ 
  &4& $s$  $\bar s$  $\nu_\tau$  $\bar\nu_\tau$ \\
\hline
NC32&24&1& $s$  $\bar s$  $u$  $\bar u$ 
  &2& $d$  $\bar d$  $c$  $\bar c$ \\
\hline
NC24&25&1& $\mu^-$  $\mu^+$  $\tau^-$  $\tau^+$ 
  &&\\
\hline
NC48&26&1& $e^-$  $e^+$  $\mu^-$  $\mu^+$ 
  &2& $e^-$  $e^+$  $\tau^-$  $\tau^+$ \\
\hline
NC48&27&1& $\mu^-$  $\mu^+$  $\mu^-$  $\mu^+$ 
  &2& $\tau^-$  $\tau^+$  $\tau^-$  $\tau^+$ \\
\hline
NC144&28&1& $e^-$  $e^+$  $e^-$  $e^+$ 
  &&\\
\hline
NC64&29&1& $d$  $\bar d$  $d$  $\bar d$ 
  &2& $s$  $\bar s$  $s$  $\bar s$ \\
\hline
NC48&30&1& $e^-$  $e^+$  $d$  $\bar d$ 
  &2& $e^-$  $e^+$  $s$  $\bar s$ \\
\hline
NC24&31&1& $\mu^-$  $\mu^+$  $d$  $\bar d$ 
  &3& $\tau^-$  $\tau^+$  $d$  $\bar d$ \\
&  &2& $\mu^-$  $\mu^+$  $s$  $\bar s$ 
  &4& $\tau^-$  $\tau^+$  $s$  $\bar s$ \\
\hline
NC32&32&1& $d$  $\bar d$  $s$  $\bar s$ 
  &&\\
\hline 
\hline
\multicolumn{6}{|c|}{} \\
\multicolumn{6}{|c|}{\bf NC+HIGGS}\\
\hline
process type& iproc& ich & final state& ich & final state \\
\hline
\hline
NC21&33&1& $b$  $\bar b$  $\nu_e$  $\bar\nu_e$ 
  &&\\
\hline
NC11&34&1& $b$  $\bar b$  $\nu_\mu$  $\bar\nu_\mu$ 
  &2& $b$  $\bar b$  $\nu_\tau$  $\bar\nu_\tau$ \\
\hline
NC33&35&1& $b$  $\bar b$  $u$  $\bar u$ 
  &2& $b$  $\bar b$  $c$  $\bar c$ \\
\hline
NC25&36&1& $b$  $\bar b$  $\mu^-$  $\mu^+$ 
  &2& $b$  $\bar b$  $\tau^-$  $\tau^+$ \\
\hline
NC50&37&1& $b$  $\bar b$  $e^-$  $e^+$ 
  && \\
\hline
NC33&38&1& $b$  $\bar b$  $d$  $\bar d$ 
  &2& $b$  $\bar b$  $s$  $\bar s$ \\
\hline
NC84&39&1& $b$ $\bar b$  $b$  $\bar b$ 
  &&\\
\hline
\multicolumn{6}{c}{ }\\
\multicolumn{6}{c}{\bf Table 3 }\\
\end {tabular}
\ec

\newpage
\begin{figure}[ht]
\vspace{0.1cm}
\centerline{
\epsfig{figure=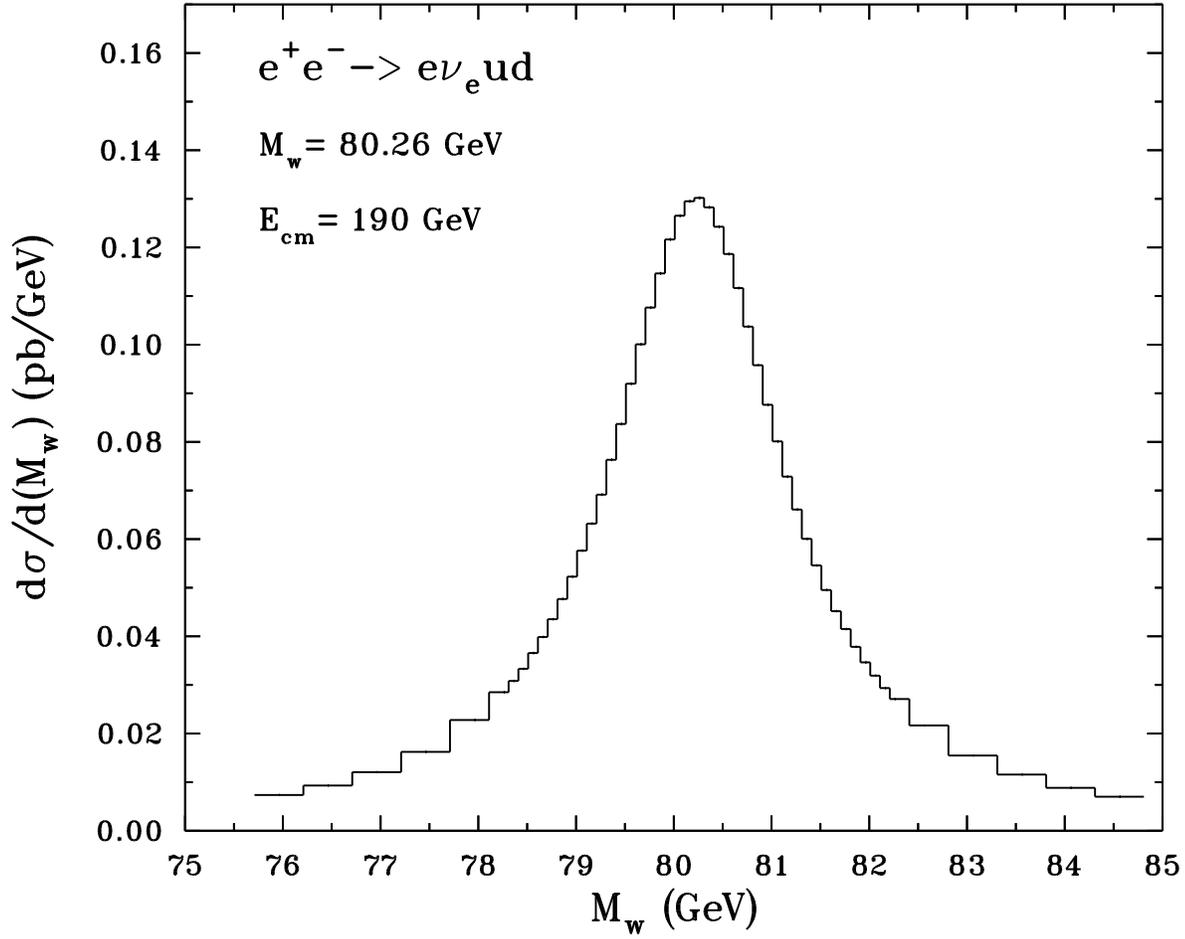,height=14cm,angle=0}
}
\caption{ Distribution of the invariant mass corresponding to $M_{W^+}$,
computed in test run 1.}
\label{fig5}
\end{figure}

\newpage
\begin{figure}[ht]
\vspace{0.1cm}
\centerline{
\epsfig{figure=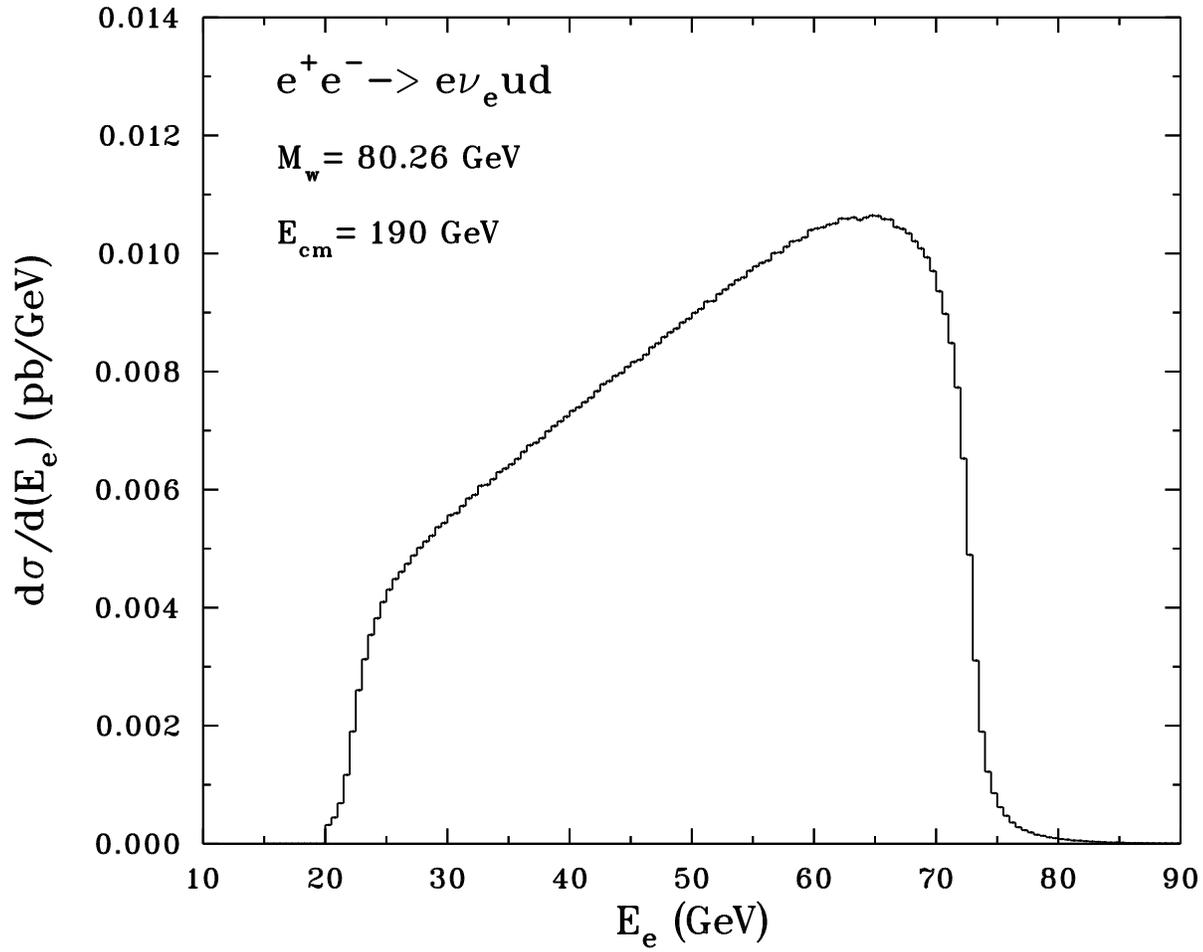,height=14cm,angle=0}
}
\caption{ Distribution of the energy of the outgoing electron computed in 
test run 1.}
\label{fig2}
\end{figure}

\end{document}